\documentclass[11pt,letterpaper,epsf,preprint,floatfix,amsmath,onecolumn]{revtex4}

\usepackage{amsmath}
\usepackage{epsfig}
\usepackage{epsf}
\usepackage{array}
\usepackage{verbatim}
\usepackage{hyperref}

\newcommand{\re}[1]{(\ref{#1})}

\newcommand{\up}{\uparrow}

\newcommand{\dn}{\downarrow}

\newcommand {\dis}{\displaystyle}

\newcommand{\beg}{\begin{equation}}
\newcommand{\en}{\end{equation}}

\newcommand{\eps}{\epsilon}
\newcommand{\lam}{\lambda}

\newcommand{\w}{\omega}

\newcommand{\eref}[1]{Eq.~(\ref{#1})}

\begin{document}

\title{Normal and anomalous solitons in the theory of dynamical Cooper pairing}
\author{Emil A. Yuzbashyan}
\affiliation{Center for Materials Theory, Department of Physics and Astronomy,
Rutgers University, Piscataway, New Jersey 08854, USA}

\begin{abstract}

We obtain multi-soliton solutions of the time-dependent
Bogoliubov-de Gennes equations or, equivalently, Gorkov equations
that describe the dynamics of a fermionic condensate in the
dissipationless regime. There are two kinds of solitons -- {\it
normal} and {\it anomalous}. At large times, normal multi-solitons
asymptote to  unstable stationary states of the BCS Hamiltonian
with zero order parameter (normal states), while the anomalous
ones tend to  eigenstates characterized by a nonzero anomalous
average. Under certain circumstances,  multi-soliton solutions
break up into sums of single solitons. In the linear analysis near
the stationary states, solitons correspond to unstable  modes.
Generally, they are nonlinear extensions of these modes, so that a
stationary state with $k$ unstable modes gives rise to a
$k$-soliton solution. We relate parameters of the multi-solitons
to those of the asymptotic stationary state, which determines the
conditions necessary for exciting solitons. We further argue that
the dynamics in many physical situations is multi-soliton.

\end{abstract}

\maketitle

\tableofcontents

\newpage

\section{Introduction and summary of the results}

Recent years have witnessed renewed experimental and theoretical interest in far from equilibrium phenomena in
strongly interacting many-body systems at low temperatures. Examples include
non-stationary Kondo and other impurity models[\citealp{Goldhaber-Gordon}--\citealp{Mehta}], quenched Luttinger
liquids[\citealp{Schumm}--\citealp{Gritsev}], electron spin dynamics
induced by hyperfine interactions[\citealp{Ono}--\citealp{Al-Hassanieh}] etc. On the theory side, there
is a considerable  effort to develop new  approaches to
nonequilibrium
 many-body physics. This presents a significant challenge as conventional techniques are often inadequate for the
 description of these phenomena.
In particular, there have been major advances in the theory of dynamical fermionic pairing in the collisionless
regime[\citealp{galaiko}--\citealp{spectroscopy}].
This problem is long  known to be accurately described  by the time-dependent
Bogoliubov-de Gennes equations, which in this case are a set of coupled nonlinear integro-differential
equations\cite{volkov,galperin,single,gensol1}. Nevertheless, it was
not until recently that this nonlinear system was realized to be exactly solvable\cite{gensol,gensol1,dicke}. The exact solution
proved to be a unique approach to
the problem of dynamical pairing and has been extensively exploited to obtain analytical information about its
key physical properties. For example, a nonequilibrium ``phase diagram'' of a homogeneous Bardeen-Cooper-Schriffer (BCS)
superfluid with a number
of novel phases, as well
as their responses to existing experimental probes were predicted analytically\cite{emilts,Levitov2006,Emil2006,spectroscopy}.

However, while much  attention was focused on the asymptotic states of the condensate at large times, the transient dynamics
has not been fully explored. Most importantly, nonlinear integrable systems are known to exhibit a remarkable class of
multi-soliton solutions that play a central role in understanding and predicting their  properties. Physical
solutions can often be represented as a  superposition of solitons making a quantitative analysis
possible. For instance,
one can show that the dynamics giving rise to nonequilibrium ``phases'' mentioned above is multi-soliton, see below.
The existence and properties of solutions of this type are also of a general interests
from the point of view of nonlinear physics due to the nonlocal nature of the  BCS problem
distinguishing it from  familiar
integrable  systems, such as nonlinear Shr\"odinger, Korteweg-de Vries, sine-Gordon etc.

In this paper we construct multi-soliton solutions to the dynamical fermionic pairing problem, see Figs.~\ref{1nsfig} -- \ref{2asfig}.
We establish a direct correspondence
between solitons and the stationary states of the mean-field BCS Hamiltonian, such that each soliton solution   asymptotes to an
eigenstate
at times $t\to\pm\infty$.  This also identifies conditions necessary for exciting solitons. There are two distinct types of solitons
-- {\it normal} and {\it anomalous}. Normal solitons asymptote to stationary states that are simultaneous eigenstates of
a Fermi gas and  the mean-field BCS Hamiltonian and are characterized by a zero anomalous average.  For anomalous solitons the
asymptotic value of this average is finite,
$\Delta(t\to\pm\infty)\ne0$. As the separation between solitons is increased, the multi-soliton solution breaks up into a
simple sum of single solitons (see e.g. Figs.~\ref{2nsfig} and \ref{3nsfig})-- one of the defining properties of solitons.

In the rest of this section, we briefly formulate the problem and then  summarize the main  results.
The collisionless dynamics of a fermionic superfluid  can be described by the Bogoliubov-de
Gennes   equations\cite{single,gensol1}
\beg
i \frac{d}{dt}
\left(\begin{array}{c}
U_m \\
V_m \\
\end{array}\right)=
\left(\begin{array}{cc}
\eps_m & \Delta(t)\\
\Delta^*(t) & -\eps_m\\
\end{array}\right)
\left(\begin{array}{c}
U_m \\
V_m \\
\end{array}\right)
\label{bdg} \en where $\Delta(t)=g\sum_m U_m V_m^*$ is the
anomalous average, $\eps_m$ are the single fermion energies
relative to the Fermi level $\eps_F$,  and $g$ is the coupling
constant. These nonlinear equations are known to be integrable for
any number of Bogoliubov amplitudes $(U_m,
V_m)$\cite{gensol,gensol1,dicke}.  In the continuum limit  the
summation in the expression for $\Delta(t)$ is replaced by
integration and Eqs.~\re{bdg} become integrable nonlinear
integro-differential equations.   Each solution of Eq.~\re{bdg}
yields a many-body wave function \beg
|\Psi(t)\rangle=\prod_m[U_m^*(t)+V_m^*(t) \hat c_{m\up}^\dagger
\hat c_{m\dn}^\dagger]|0\rangle, \label{wf} \en where the product
is taken only over unblocked levels -- levels that are either
unoccupied or doubly occupied.

 As we  demonstrate in subsequent sections, solitons tend to unstable stationary states of the mean-field BCS Hamiltonian at
 $t\to\pm\infty$.
 The mean-field Hamiltonian has two types of eigenstates -- {\it normal} and {\it anomalous}. Anomalous stationary states have a nonzero
 constant value $\Delta_a$ of the anomalous average and are solutions of Eq.~\re{bdg} of the form\cite{BCS,Tinkham}
$(U_m, V_m)=(U^0_m, V^0_m) e^{-iE_m t}$, where $(U^0_m,V^0_m)$ and $E_m$ are the eigenvectors and eigenvalues of the $2\times 2$ matrix
on the right hand side of Eq.~\re{bdg}. There are two states $E_m=\pm \sqrt{\eps_m^2 +\Delta^2}$ for each $\eps_m$.   The BCS ground
state has $E_m<0$ for all $m$. A state where $E_r>0$ while $E_{m\ne r}<0$ describes a single excited
pair\cite{BCS,ander1} and has energy
$2\sqrt{\eps_r^2+\Delta^2}$ above the ground state. We note also that an excited pair introduces a discontinuity in the average fermion
occupation number $n(\eps_m)=1-\eps_m/E_m$, since $E_m$ changes sign at $m=r$. Normal eigenstates have $\Delta=0$ and amplitudes
$(U_m, V_m)$ equal to either $(0, e^{i\eps_m t})$ or $(e^{-i\eps_m t}, 0)$.  Both types of
eigenstates are exact stationary states of the mean-field BCS Hamiltonian
\beg
\hat H=\sum_{ j; \sigma=\dn, \up}\eps_{j} \hat c_{j \sigma }^\dagger \hat c_{j \sigma}-\sum_{j, k}\left(\Delta \hat c_{j\up}^\dagger
\hat c_{j\dn}^\dagger +\mbox{h.c.}\right)  ,
\label{bcs1}
\en
where $\Delta=g\sum_k \langle \hat c_{ k\dn} \hat c_{ k\up}\rangle$, and $\hat c_{j \sigma }^\dagger$ and $\hat c_{j \sigma }$ are the
creation and annihilation operators for the two fermion species. Normal states are also eigenstates of the Fermi gas -- the first term
in Eq.~\re{bcs1}. For example, the Fermi ground state  is a normal eigenstate with $U_m=0$ for $\eps_m<0$ and $V_m=0$ for $\eps_m>0$,
i.e. all single particle states below the Fermi level  occupied and states above it  empty.

A linear analysis of Eq.~\re{bdg} around  stationary states shows that some of
them are  unstable\cite{Elihu,single}. These states give rise to solitons, as is  typical in integrable nonlinear dynamics. For example,
a simple pendulum displays a soliton solution when started in its unstable equilibrium with zero velocity.
In the phase space the soliton is the
separatrix connecting the unstable equilibrium to itself. The same is true for example for the single soliton solution of the
Korteweg-de Vries
equation\cite{Arnold}.
Similarly in the present case   the unstable modes start to grow exponentially and become solitons due to nonlinear effects.
In a certain sense,
solitons can be viewed as nonlinear extensions of the unstable modes.

Now let us summarize soliton solutions of the Bogoliubov-de Gennes equations \re{bdg}. Detailed derivation and discussion of these as
well as some other solutions can be found in subsequent sections.  Here we present only the results for the amplitude of the order
parameter $|\Delta(t)|$.

\subsection{Normal solitons}

\begin{figure}
\includegraphics[width=0.45\textwidth]{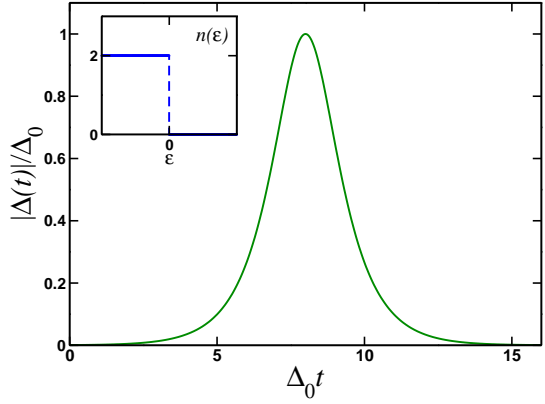}
\caption{(color online) $k=1$ normal-soliton solution of the Bogoliubov-de Gennes equations \re{bdg} \cite{single}, where $\Delta_0$
is the ground state BCS gap. At $t\to\pm\infty$ the solution asymptotes to the Fermi ground state. The inset
shows the average fermion occupation number $n(\eps)$ in this state. A single discontinuity ($2k-1=1$) in $n(\eps)$
at the Fermi energy gives rise to a single soliton, see the text.}
\label{1nsfig}
\end{figure}

First, we present normal multi-solitons derived in Sec.~\ref{normal}.
The 1-soliton solution of a normal type has been previously obtained in Ref~\onlinecite{single}. The amplitude of
the order parameter
 has the following form:
\beg
|\Delta(t)|=\frac{2\gamma}{\cosh(2\gamma t+\alpha)}
\label{1ns},
\en
where $\alpha$ is a real parameter.
At $t\to\pm\infty$ the corresponding wave function \re{wf} asymptotes to the Fermi ground state. The latter is unstable in the presence
of the pairing interaction. Indeed, a linear analysis of equations of motion \re{bdg} around the Fermi ground state shows a single
unstable normal mode, which grows as $|\Delta(t)|\propto e^{2\gamma t}$ \cite{Elihu,single} as can  be seen from Eq.~\re{1ns} in the
$t\to-\infty$
limit. The plot of Eq.~\re{1ns} is a single peak centered at $t=-\alpha/2\gamma$, see Fig.~\ref{1nsfig}.  Its height (the amplitude of
the soliton) and  width are controlled
by the parameter $2\gamma$. In the present case $2\gamma =\Delta_0$, where $\Delta_0$ is the ground state BCS gap. This parameter
can  be interpreted as an imaginary ``frequency'' of the unstable normal mode, while Eq.~\re{1ns}  as
an extension of this mode to the nonlinear regime. Below we will see that the number of unstable modes and consequently the
number of solitons   corresponding to a given stationary state is related
to the number of discontinuities in the  average fermion occupation number $n(\eps)$ in this state. Specifically,
$2k-1$ discontinuities (jumps) in $n(\eps)$ lead to up to $k$ coupled normal solitons. The Fermi ground state has a single discontinuity
at the Fermi level, giving rise to the 1-soliton solution.

\begin{figure}
{\bf a)}\includegraphics[width=0.45\textwidth]{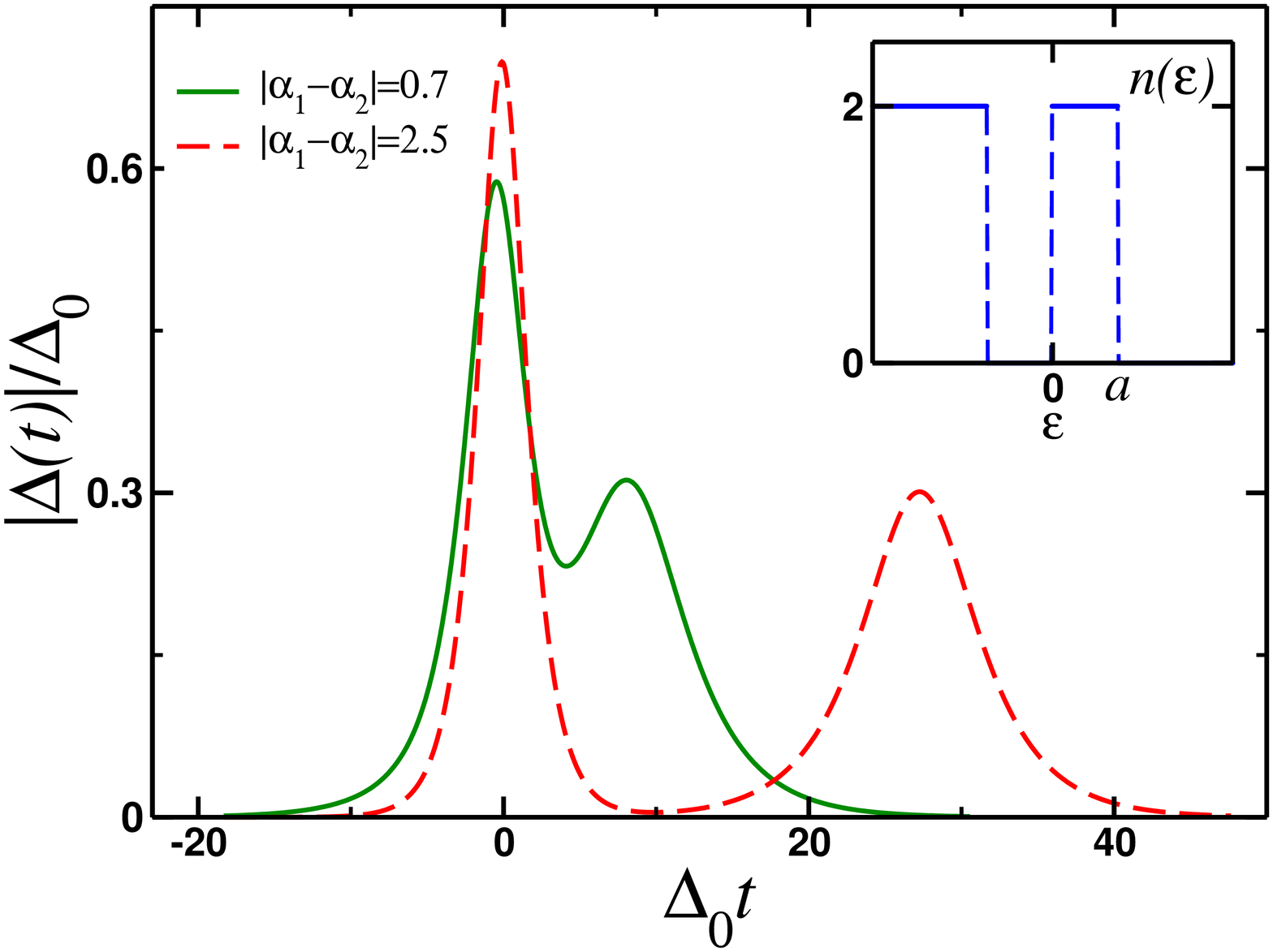}
{\bf b)}\includegraphics[width=0.45\textwidth]{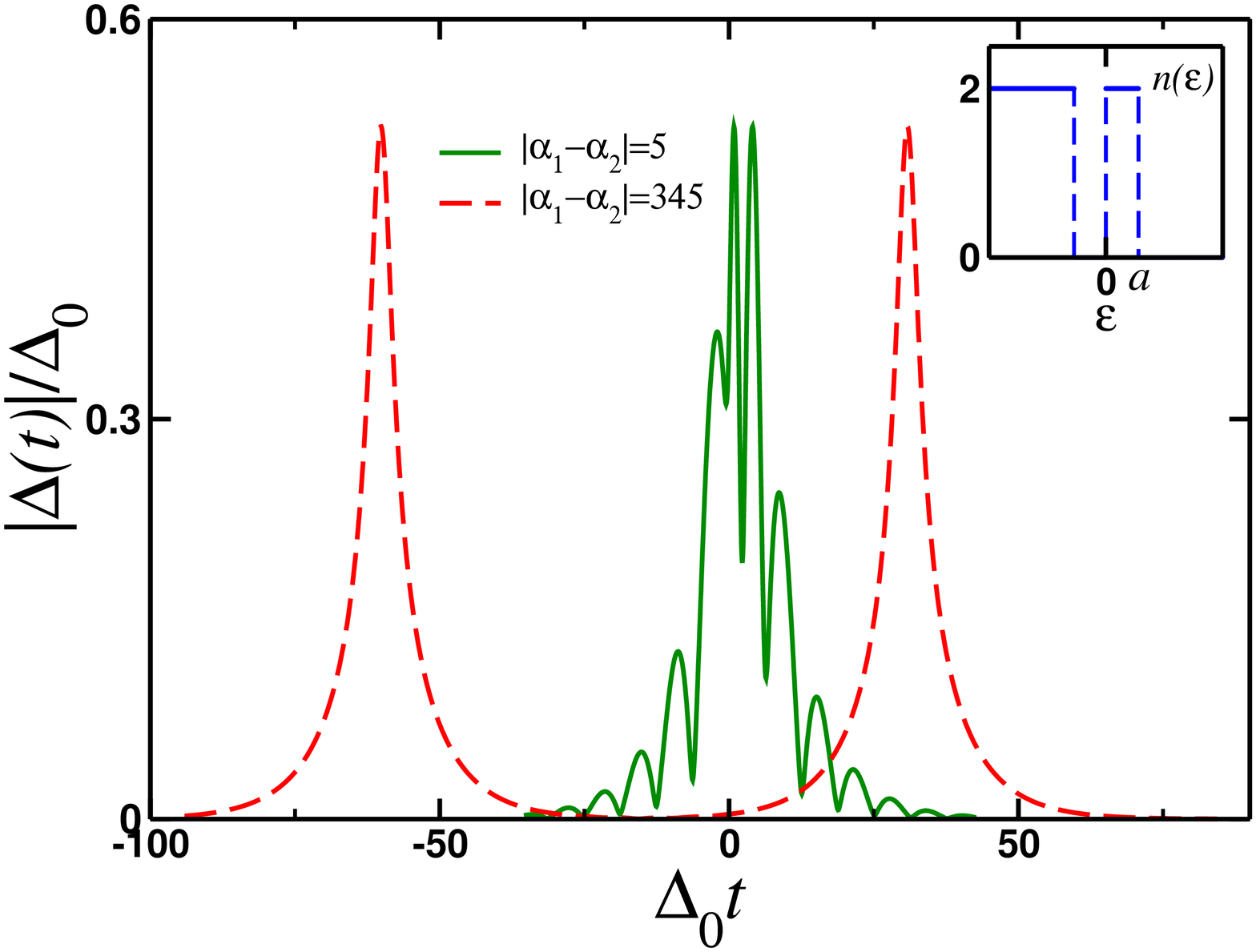}
\caption{(color online) $k=2$ normal soliton solutions  to the Bogoliubov-de Gennes equations \re{bdg}. $\Delta_0$
is the ground state BCS gap. At $t\to\pm\infty$ the 2-solitons tend to a normal state characterized by
$2k-1=3$ jumps at $\eps=0$ and $\pm a$ in the average fermion occupation number $n(\eps)$ (insets).
 a) is obtained from \eref{h1}, where $\gamma_{1,2}$ are related to $a$ and $\Delta_0$ by \eref{44}, and
  $4.25 a=\Delta_0$. b) corresponds to \eref{h2intro} with $4\mu=\Delta_0$,
  $4\gamma=\sqrt{16a^2-\Delta_0^2}$ (see the text below \eref{44}), and $a=3.75\Delta_0$. $\phi_{1,2}=0$ in both cases. For large
  separation, $|\alpha_1-\alpha_2|\gg 1$, the 2-soliton solutions split into  two single solitons (dashed lines), see
  Eqs.~\re{1ns}, \re{sumns}, \re{2nslargesep},  and Fig.~\ref{1nsfig}. For  $|\alpha_1-\alpha_2|\to 0$ the two solitons
  merge into a single peak. In b) the amplitude of this peak is modulated with a frequency $\w \simeq 4\mu=\Delta_0$ as
  the two terms in \eref{h2intro} ``rotate'' with respect to each other. }
  \label{2nsfig}
\end{figure}

The 2-normal-solitons have considerably richer structure. Let us give two examples. Both are described by
\beg
|\Delta(t)|=A\left| \frac{h(t)}{h(t)\ddot h(t) -\dot h^2(t)}\right|, \label{2ns}
\en
with different choices for the amplitude  $A$ and function $h(t)$. One choice is $A=4|\gamma_2^2-\gamma_1^2|$ and
\beg
 h(t)=e^{i\phi_1}\frac{\cosh(2\gamma_1t+\alpha_1)}{2\gamma_1}+e^{i\phi_2}\frac{\cosh(2\gamma_2t+\alpha_2)}{2\gamma_2}.
\label{h1} \en Examples of $|\Delta(t)|$ for this case are plotted
in Fig.~\ref{2nsfig}a. The second option is
$A=16\mu\sqrt{\mu^2+\gamma^2}$ and \beg h(t)=e^{-2i\mu
t+i\phi_1}\frac{\cosh (2\gamma t+\alpha_1-i\beta)}{2\gamma}+
e^{2i\mu t+i\phi_2}\frac{\cosh (2\gamma
t+\alpha_2+i\beta)}{2\gamma}, \label{h2intro} \en where $\beta$ is
the phase of $\mu+i\gamma$, i.e.
$\mu+i\gamma=\sqrt{\mu^2+\gamma^2}e^{i\beta}$. Fig.~\ref{2nsfig}b
shows graphs of $|\Delta(t)|$ obtained using \eref{h2intro}. In
both cases the wave function asymptotes  to a normal eigenstate at
$t\to\pm\infty$. This eigenstate is obtained from the Fermi ground
state by moving all fermions in the energy interval $-a<\eps_m<0$
below the Fermi level to the symmetric interval $0<\eps_m<a$ above
it. Eqs.~\re{h1} and \re{h2intro} correspond to different values
of $a$.   Note that this normal eigenstate has $2k-1=3$
discontinuities at $\eps=-a$, 0, and $a$ in $n(\eps)$ (see the
insets in Fig.~\ref{2nsfig}) thus leading to $k=2$ solitons. A
linear analysis around this state yields two unstable modes that
exponentially grow with rates $2\gamma_{1,2}$ in the first case
and $2\gamma\pm 2i\mu$ in the second. Which of the two cases is
realized depends on the ratio $a/\Delta_0$. The values of
$\gamma_{1,2}$ or $\gamma\pm i\mu$ are also fixed by $a$ and
$\Delta_0$, while  $\alpha_{1,2}$ and $\phi_{1,2}$ are arbitrary
real parameters. They control the separation between solitons and
their relative phase in the 2-soliton solution, respectively. For
sufficiently large $|\alpha_1-\alpha_2|$ Eq.~\re{2ns} always
breaks up into a sum of two single solitons of the form \re{1ns}.
The graph of $|\Delta(t)|$ in this case displays two well
separated peaks, each representing a single soliton, see
Fig.~\ref{2nsfig}.

\begin{figure}
{\bf a)}\includegraphics[width=0.45\textwidth]{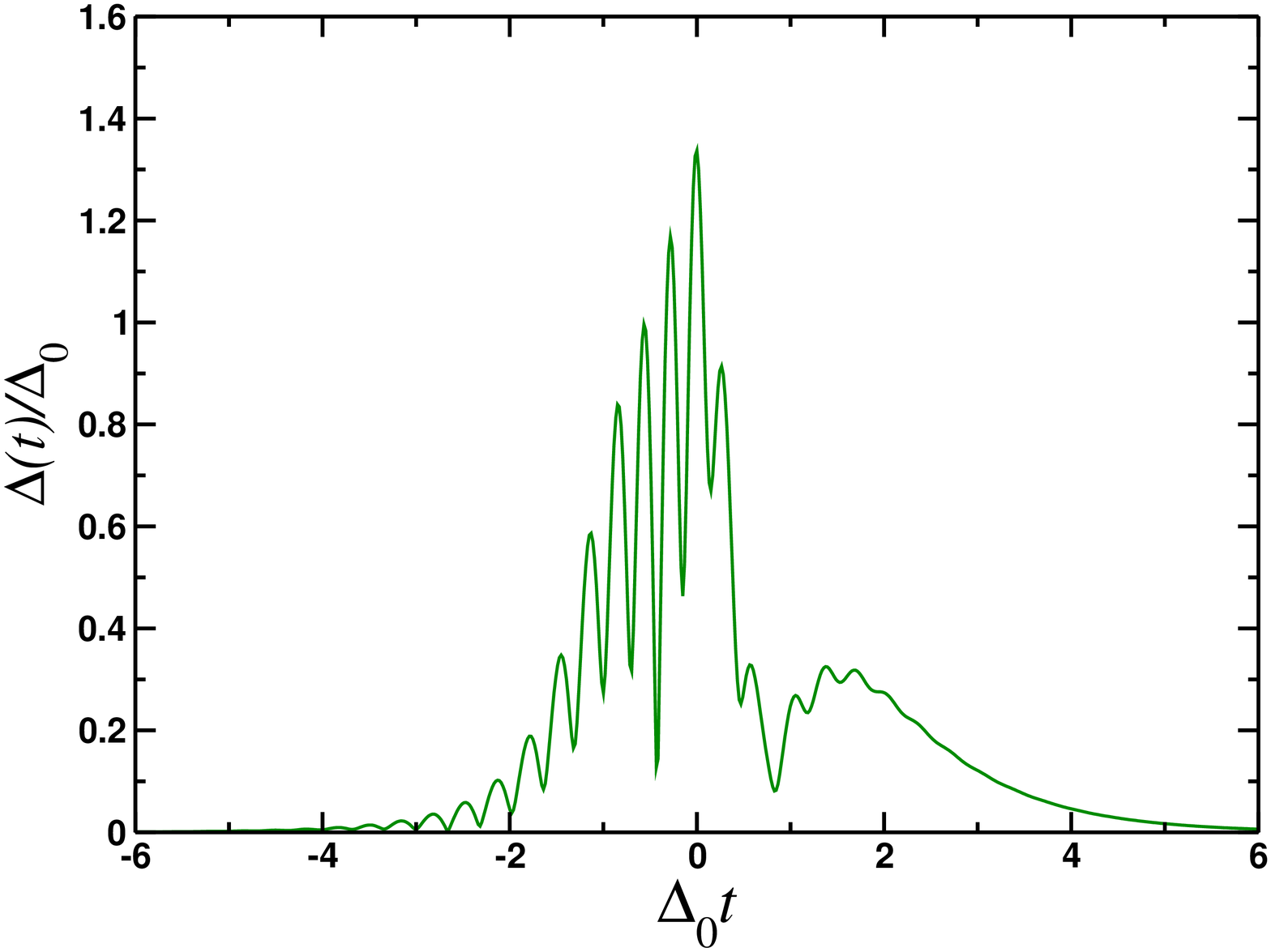} {\bf
b)}\includegraphics[width=0.45\textwidth]{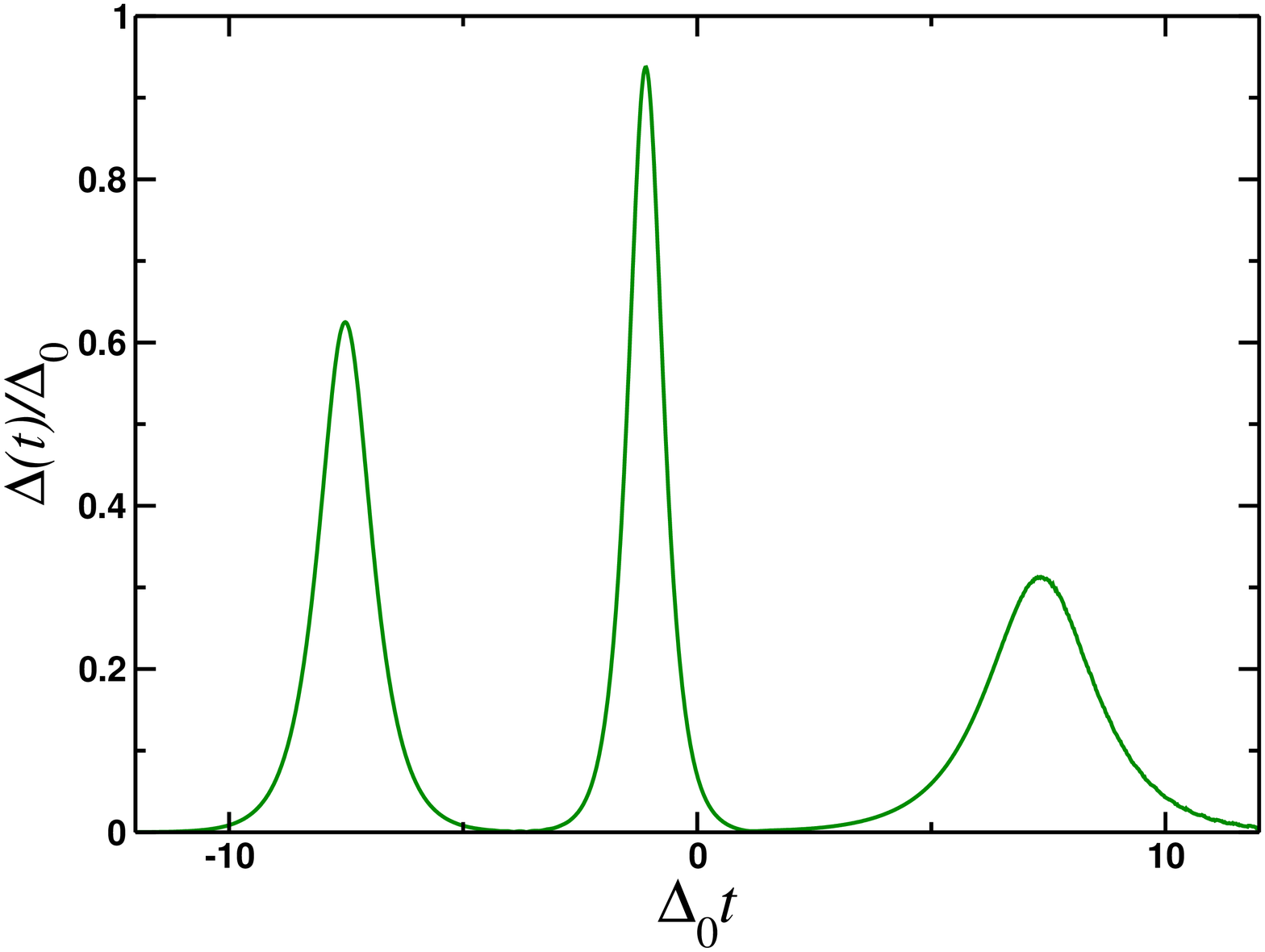}
\caption{(color online) 3-normal-soliton solution  to the
Bogoliubov-de Gennes equations \re{bdg} obtained from \eref{kns}
with random  $c_j$ and $\phi_j$. $\Delta_0$ is the ground state
BCS gap. a) and b) differ only in the values of $\alpha_j$. In a)
$\alpha_2-\alpha_1=1.00$ and $\alpha_3-\alpha_2=1.25$; in b)
$\alpha_2-\alpha_1=10$ and $\alpha_3-\alpha_2=12$. We see that for
large differences between $\alpha_j$ in b) the 3-soliton breaks up
into a sum of three well separated individual solitons, see
\eref{sumns}, while in a) the same solution but with small
differences describes a complicated interference  between the
three solitons.} \label{3nsfig}
\end{figure}

The general $k$-soliton solution of the normal type has the form (see Sec.~\ref{gennormal})
\beg
|\Delta(t)|=2^{2k-1}\left| \frac{D_{k-1}}{D_k}\right|,
\label{kns}
\en
where $D_r$ is the following determinant:
\beg
D_r=
\left|
\begin{array}{lll}
f & \dots & f^{(r-1)}\\
\vdots & & \vdots\\
f^{(r-1)}&\dots & f^{2(r-1)}\\
\end{array}
\right|,
\label{gapnormal}
\en
$f^{(m)}$ is the $m$th derivative of the function $f(t)$ with respect to $t$, and
\beg
f(t)=\sum_{j=1}^{2k} \frac{ A(c_j) e^{-2i c_j t} }{\prod_{m\ne j}(c_j-c_m)},
\label{f}
\en
The set of $2k$ complex parameters $c_m$ (``frequencies'' of the unstable modes) is complex conjugate to itself. Let us order this set
 so that $c_{k+l}=c_l^*$ and $\mbox{Im}(c_l)>0$ for
$l=1,\dots,k$. The constants $A(c_l)$ and $A(c_{k+l})$
are related as follows
\beg
A(c_l)=e^{\alpha_l+i\phi_l},\quad A(c_{k+l})=-e^{-\alpha_l+i\phi_l},
\label{A}
\en
where $\alpha_l$ and $\phi_l$ are arbitrary real parameters. The  single soliton \re{1ns} is obtained
from Eq.~\re{kns} by setting $k=1$ and $c_1=\mu+i\gamma$. The 2-soliton corresponds to  $k=2$
  and $c_{1,2}=i\gamma_{1,2}$ or $c_{1,2}=i\gamma\pm\mu$ to get Eq.~\re{h1} or \re{h2intro}, respectively.
See also Fig.~\ref{3nsfig} for examples of 3-normal-solitons.

At $t\to\pm\infty$ the $k$-normal-soliton tends to a normal eigenstate that has at least $2k-1$ discontinuities in the distribution
function $n(\eps)$. Linearizing the Bogoliubov-de Gennes equations around this state at large negative $t$, one obtains $k$ unstable
normal modes that grow as $e^{-2ic_lt}$ for $l=1,\dots,k$. When the differences between parameters $\alpha_l$ are large the $k$-soliton
solution \re{kns} breaks up into a sum of $k$ single solitons,
\beg
|\Delta^{(k)}(t, \{c_i,\alpha_i,\phi_i\})|\approx\sum_{i=1}^k |\Delta^{(1)}(t, \mbox{Im} (c_i),\alpha_i)|,
\label{sumns}
\en
where $\Delta^{(k)}(t)$ denotes the $k$-normal-soliton  \re{kns} and $|\Delta^{(1)}(t, \mbox{Im} (c_i),\alpha_i)|$
stands for the single soliton \re{1ns} with $\gamma=\mbox{Im} (c_i)$ and $\alpha=\alpha_i$. In this case, the plot of $ |\Delta(t)|$
shows
$k$ well separated
peaks (individual solitons) as illustrated in Figs.~\ref{2nsfig} and \ref{3nsfig}.
 $\mbox{Im}(2c_l)$ set the amplitudes  and widths of individual solitons, $\mbox{Re}(2c_l)$ are the frequencies with which they
``rotate'' with respect to one another as in Eq.~\re{h2intro}, where $\mbox{Re}(c_{1,2})=\pm\mu$, and $\alpha_l$ and $\phi_l$ determine
the separation between the solitons and their relative phases, respectively.

\subsection{Anomalous solitons}

\begin{figure}
\includegraphics[width=0.45\textwidth]{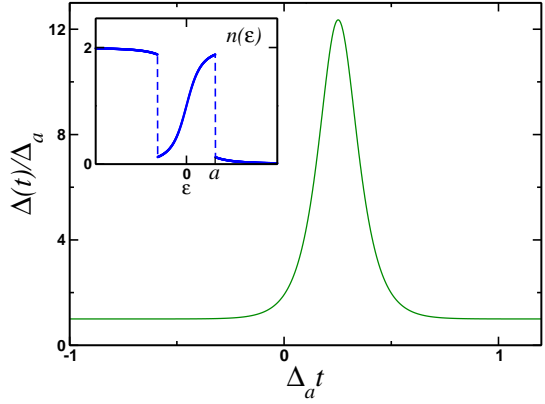}
\caption{(color online) $k=1$ anomalous soliton solution to the Bogoliubov-de Gennes equations \re{bdg} as given by
 \eref{1as}. At $t\to\pm\infty$ the system tends to an eigenstate of the BCS Hamiltonian \re{bcs1} with order parameter
 $\Delta_a$. This eigenstate is characterized by $2k=2$ jumps at $\eps=\pm a$
 in the average fermion occupation number $n(\eps)$ (inset). $\gamma$ and $\Delta_a$ in \eref{1as} are related to
 $a$ and the ground state gap $\Delta_0$ via Eqs.~\re{1g} and \re{dela}. Here $a=1.47\Delta_a$ and $\Delta_a=.09\Delta_0$.}
\label{1ansfig}
\end{figure}

Next, let us summarize the results of Sec.~\ref{anomalous} for anomalous solitons.
 A single anomalous soliton has the following form (see also Fig.~\ref{1ansfig}):
\beg
\Delta(t)-\Delta_a=\frac{2(\gamma^2-\Delta_a^2)}{\Delta_a\pm \gamma \cosh( \lam t+\alpha)},
\label{1as}
\en
where $\lam=2\sqrt{\gamma^2-\Delta_a^2}$ and $\alpha$ is an arbitrary real parameter as before.
As $t\to\pm\infty$ the state of the system tends to  an anomalous eigenstate with the value of the BCS order parameter equal to
$\Delta_a$. In this state, all pairs in a certain energy interval around the Fermi level are excited, i.e.
$E_m=\sqrt{\eps_m^2+\Delta_a^2}$ for $|\eps_m|<a$. As a result, the distribution function $n(\eps)$ has two jumps
at $\eps=\pm a$ (inset in Fig.~\ref{1ansfig}).  A linear analysis around this anomalous eigenstate   shows a single unstable mode that grows
as $e^{\lam t}$. In general,  $2k$ jumps in the distribution function of a stationary anomalous state   give rise to up to
$k$ anomalous solitons. Note also that the anomalous soliton \re{1as} generalizes the normal one \re{1ns} and turns into it
when $\Delta_a=0$.

\begin{figure}
{\bf a)}\includegraphics[width=0.45\textwidth]{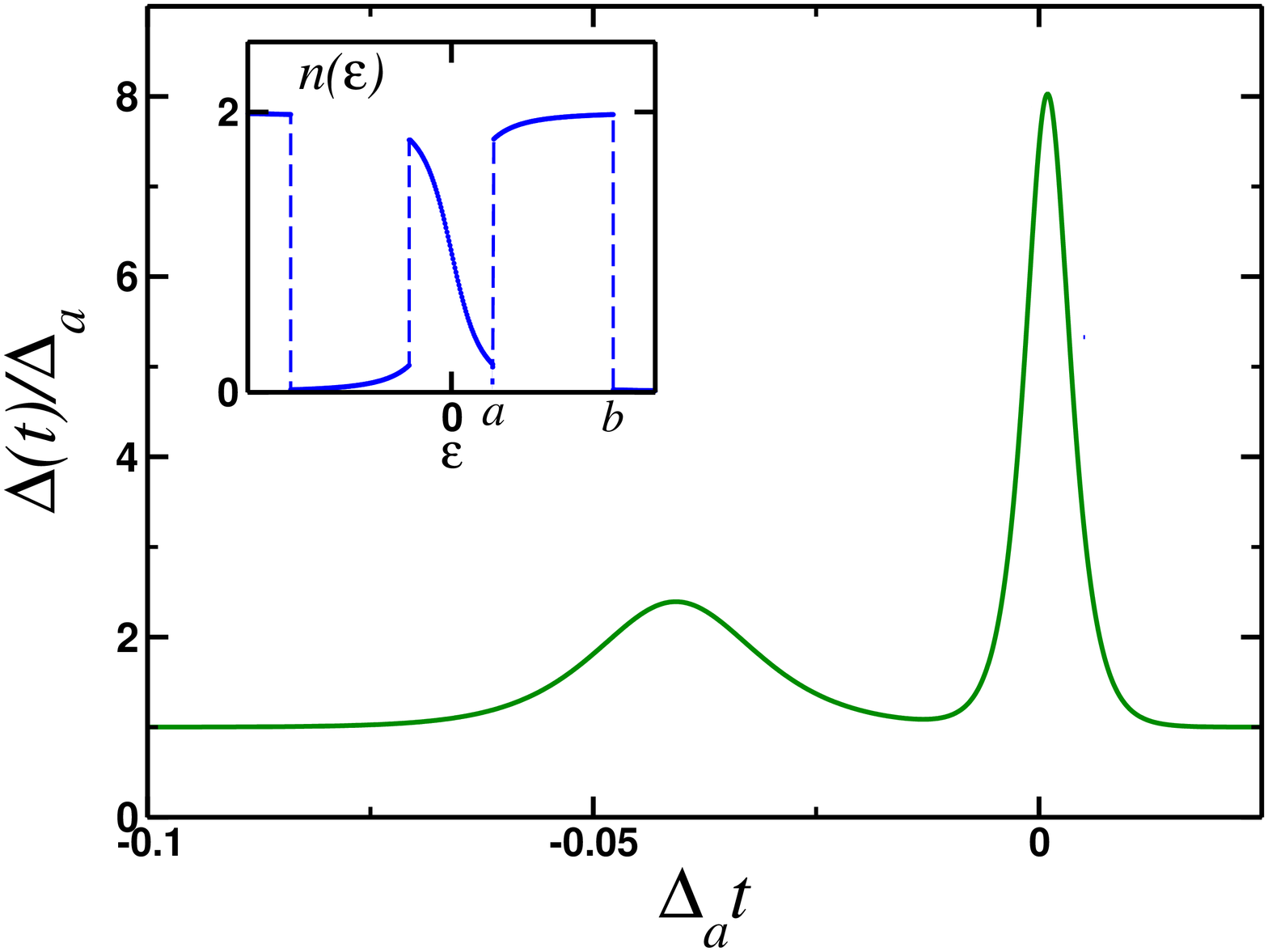}
{\bf b)}\includegraphics[width=0.45\textwidth]{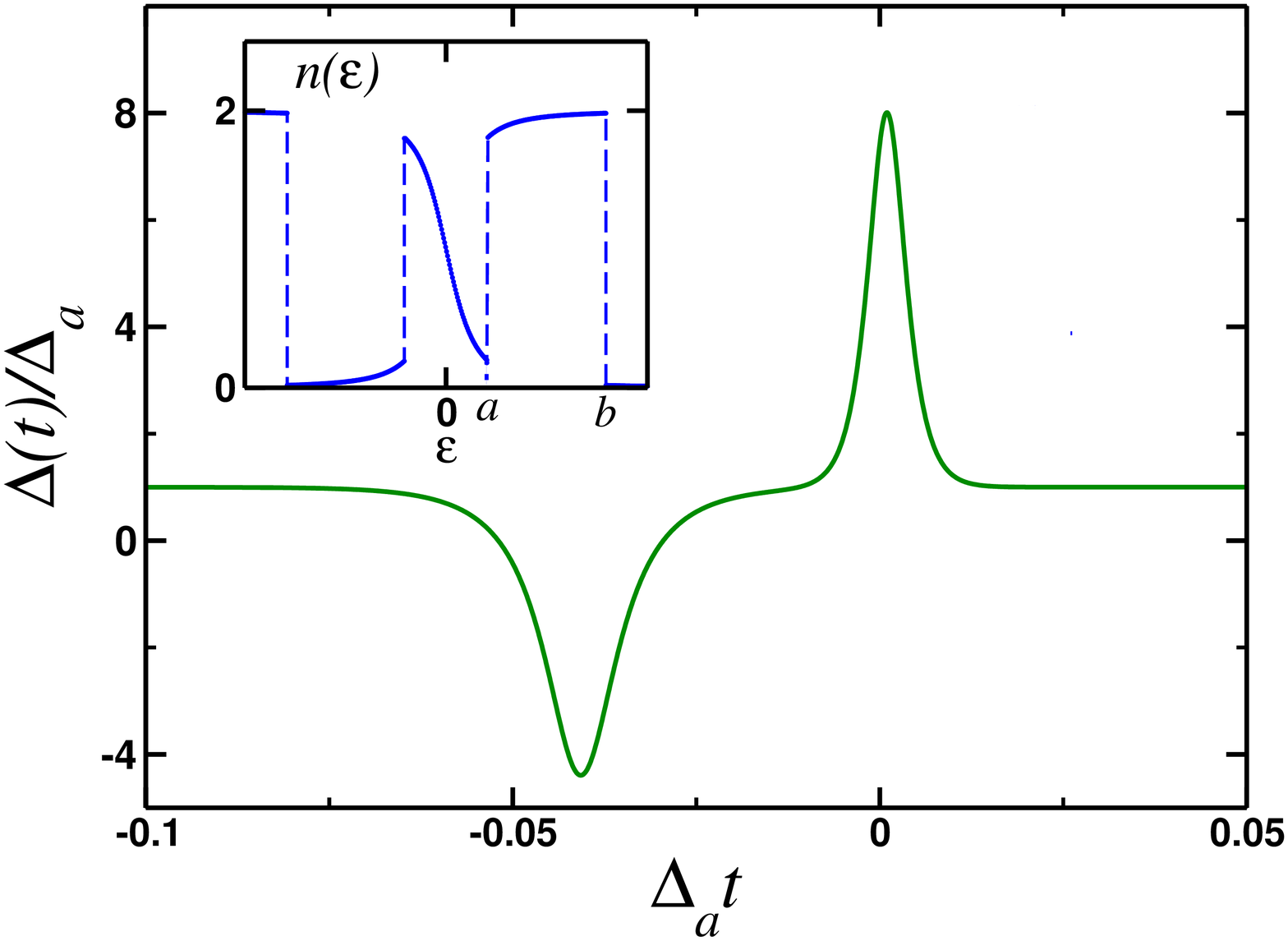}
\caption{(color online) $k=2$ anomalous soliton solutions of the Bogoliubov-de Gennes equations \re{bdg}, see
 Eqs.~\re{2as} and \re{han}. a) corresponds to the choice of signs $++$ and b) to $-+$ in \eref{han}.
  At $t\to\pm\infty$ both 2-anomalous-solitons tend to the same stationary state of the mean-field
  BCS Hamiltonian. In this state, $n(\eps)$ has $2k=4$ discontinuities at $\eps=\pm a$ and $\pm b$ (insets) and the
  anomalous average is equal to $\Delta_a$. Parameters $\gamma_{1,2}$ in \eref{han} and the value of $\Delta_a$
  are determined by $a, b$, and the ground state gap $\Delta_0$, see Eqs.~\re{2g} and~\re{56}. For the above plots
  we used $a=.87\Delta_a$, $b=4.27\Delta_a$, and $16\Delta_a=\Delta_0$.}
  \label{2asfig}
\end{figure}

A general $k$-anomalous-soliton can also
be constructed within our approach. However,  here we present only an example of a 2-anomalous-soliton solution
\beg
\Delta(t)-\Delta_a=\frac{h(t)}{2\left[h(t)\ddot h(t) -\dot h^2(t)\right]},
\label{2as}
\en
where
\beg
h(t)=\frac{\Delta_a}{\lam_1^2\lam_2^2}\pm \frac{\gamma_1 \cosh (\lam_1 t+\alpha_1)}{\lam_1^2 (\lam_1^2-\lam_2^2)}\pm \frac{\gamma_2
\cosh (\lam_2 t+\alpha_2)}{\lam_2^2 (\lam_2^2-\lam_1^2)},
\label{han}
\en
The $\pm$ signs can be chosen independently of each other.
As in the case of the {1-anomalous-soliton}, at large times the wave function asymptotes to an anomalous stationary state
with order parameter $\Delta_a$. This state has
two unstable modes that grow exponentially with rates $\lam_{1,2}=2\sqrt{\gamma_{1,2}^2 -\Delta_a^2}$. The graph of the
2-anomalous-soliton solution \re{2as} consists of two peaks or dips depending on the choice of signs in Eq.~\re{han}, see
Fig.~\ref{2asfig}.
The parameters $\alpha_{1,2}$ take arbitrary real values. Their difference, $|\alpha_1-\alpha_2|$, determines the separation
between the peaks in time.
Similarly to \eref{sumns}, at large separations   the $k$-anomalous-soliton turns into a sum of individual solitons of the
form~\re{1as}
\beg
  \Delta^{(k)}(t, \{\lam_i, \alpha_i\}) -\Delta_a\approx\sum_{i=1}^k \left[\Delta^{(1)}(t, \lam_i,\alpha_i)-\Delta_a\right],
\label{sumas}
\en
where $\Delta^{(1)}(t, \lam_i,\alpha_i)$ is the single anomalous soliton \re{1as} with $\lam=\lam_i$ and $\alpha=\alpha_i$.

The rest of the paper is organized as follows: in the next section, we review the basic setup of the problem and the tools (Lax
vector and separation variables) necessary for deriving solitons. In Sec.~\ref{normal}, we perform linear analysis of equations of
motion around normal and anomalous stationary states. This section also provides examples of normal and anomalous eigenstates that
give rise to one and two normal and anomalous solitons. Sections \ref{normal} and \ref{anomalous} are devoted to a detailed
derivation of soliton solutions and a discussion of their main properties.

\section{Review of the basic setup and relevant previous results}
\label{review}

In this section we discuss the basic setup of the problem and introduce our notation
(see Refs.~\onlinecite{single,gensol1} for more details). We also review the
properties of the
exact solution\cite{gensol,gensol1,dicke,emilts}
of the equations of motion needed for obtaining and analyzing the multi-soliton solutions summarized in the
previous section.

\subsection{Notations and basic equations}

Here we review the model Hamiltonian and its mean-field equations of motion \re{bdg}. The latter can be reformulated as equations
of motion for classical spins (angular momenta) -- this is the form we will be primarily using.   We also describe normal and
anomalous stationary states in terms of the spin variables.

The dissipationless dynamics of a fermionic superfluid  can be
modeled by the reduced BCS
Hamiltonian\cite{volkov,galperin,single,BCS}
\beg
\hat H=\sum_{
j}\eps_{j} \hat n_j  -g\sum_{j, k} \hat c_{j\up}^\dagger \hat
c_{j\dn}^\dagger \hat c_{ k\dn} \hat c_{ k\up},
\label{bcs}
\en
where $\hat n_j=\sum_{ \sigma=\dn, \up}  \hat c_{j \sigma
}^\dagger \hat c_{j \sigma}$. This description is valid in the
weak coupling regime at times shorter than the energy relaxation
time $\tau_\eps$ and for a system of size $L$ smaller than the BCS
coherence length $\xi$. Under these conditions, the BCS order
parameter   is uniform in space, the interaction matrix elements
can be evaluated at the Fermi energy $\eps_F$ yielding a single
coupling constant $g$ that is independent of $j$ and $k$. The
summations in \eref{bcs} over $j$ and $k$  are restricted to
single particle energies $|\eps_j|<D$ and $|\eps_k|<D$, where $D$
is an ultraviolet cutoff for the pairing interaction. In metallic
superconductors $D\simeq \w_D$, where $\w_D$ is the Debye
frequency. For atomic fermions $D\simeq \eps_F$. The offdiagonal
interactions -- terms of the form $c_{j\up}^\dagger \hat
c_{l\dn}^\dagger \hat c_{ k\dn} \hat c_{ r\up}$ with  $l\ne j$ or
$r\ne k$ can be neglected, since they are relevant only at times
$t>\tau_\eps$.

The validity of the mean-field approach is rooted in the fact that each pair creation operator
$\hat c_{j\up}^\dagger \hat c_{j\dn}^\dagger$ in Eq.~\re{bcs} interacts with the collective pairing field
$g\sum_j   \hat c_{ k\dn} \hat c_{ k\up} $, which is expected to deviate little from its quantum mechanical average $\Delta(t)$.
For example, the mean-field is known to be exact for
the description of the low-energy properties of the Hamiltonian \re{bcs} in the limit
$\delta/\Delta_0\to0$ \cite{ander1,rich,ander}, where $\delta=\langle
\eps_{m+1}-\eps_m\rangle$ and $\Delta_0$ are the mean spacing between the single particle levels $\eps_m$ and the ground state gap,
respectively. Note that the
conditions $\delta\ll\Delta_0$ and $L\ll\xi$ are compatible in the weak coupling regime.

We are interested in solving the Heisenberg equations of motion for Hamiltonian \re{bcs} to determine the evolution of various
correlators, e.g.  $\langle\hat n_m(t)\rangle$, $\langle\hat c_{m\dn}(t) \hat c_{m\up}(t)\rangle$, and
$\langle\hat c_{m\up}^\dagger(t) \hat c_{m\dn}^\dagger(t)\rangle$.  In mean-field approach,
we replace the operator $g\sum_j   \hat c_{ k\dn}(t) \hat c_{ k\up}(t) $ in the Heisenberg equations with its quantum mechanical average
\beg
\Delta(t)=g\sum_k \langle \hat c_{ k\dn}(t) \hat c_{ k\up}(t)\rangle.
\label{deltadef}
\en
Further, introducing
\beg
\begin{array}{l}
2s_m^z=  \langle\hat n_m\rangle-1 ,\\
\\
s_m^-\equiv s_m^x- is_m^y=\langle\hat c_{m\dn} \hat c_{m\up}\rangle ,\\
\end{array}
\label{bog}
\en
we obtain\cite{ander1}
\begin{equation}
\dot{\mathbf{s}}_{m}=\mathbf{b}_m\times\mathbf{s}_{m},\quad\mathbf{b}_m=
\left(-2\Delta_{x},-2\Delta_{y},2\epsilon_{m}\right),
\label{eqsm}
\end{equation}
where $\Delta_x$ and $-\Delta_y$ are the real and imaginary parts of $\Delta=g\sum_m s_m^-$. In terms of $F_m(t)=2is_m^z$
and $G_m(t)=2i s_m^-$, Green's functions at coinciding times, Eqs.~\re{eqsm} are well-known Gorkov equations\cite{Gorkov,volkov}.
The above procedure leading to Gorkov equations is
essentially equivalent to taking the
time-dependent wave function of the system to have the product form \re{wf} at all times. Then,  the
Schr\"odinger equation takes
the form of the Bogoliubov-de Gennes equations \re{bdg}, which are in turn equivalent
to Eq.~\re{eqsm} with
\beg
2s_m^z= |V_m|^2-|U_m|^2,\quad s_m^-=U_m V^*_m.
\label{bgsp}
\en

Equations of motion \re{eqsm} are Hamilton's equations for the following classical spin (interacting angular momentum) model:
\beg
H_{cl}=\sum_{j=1}^n 2\eps_j s_j^z-g\sum_{j,k=1}^n s_j^+s_k^-,
\label{hcl}
\en
with the usual angular momentum Poisson brackets $\{s_j^x, s_j^y\}=-s_j^z$ etc. The summations in \eref{hcl}
are restricted to the subspace of unblocked (with occupation numbers $n_j=0,2$) single fermion levels $\eps_j$.
The Hamiltonian \re{bcs} does not have matrix elements  connecting the unblocked levels to blocked
ones ($n_j=1$). The latter are decoupled and their occupation numbers are conserved by the evolution.
 Note also that Eq.~\re{bdg} conserves the norm $|U_m|^2+|V_m|^2=1$ and therefore the length of the spins is fixed
$|{\bf s}_m|=1/2$.

The normal and anomalous eigenstates of the BCS Hamiltonian discussed in the Introduction correspond to equilibrium spin configurations
where each spin ${\bf s}_m$
is either parallel or antiparallel to the field ${\bf b}_m$\cite{ander1}. According to Eq.~\re{bgsp}, every such arrangement of spins
uniquely determines an eigenstate of the form
\re{wf}  and vice versa.  The anomalous eigenstates yield
\beg
2s_m^z=-\frac{e_m\eps_m}{ \sqrt{\eps_m^2+\Delta_a^2} }, \quad 2s_m^x=-\frac{e_m\Delta_a}{ \sqrt{\eps_m^2+\Delta_a^2} },
\quad e_m=\pm 1,
\label{anom}
\en
where the $x$ axis has been chosen so that the stationary
value of the order parameter, $\Delta_a$, is real. The factor $e_m=-1$ if the spin is parallel to the field
and $e_m=1$ otherwise.
The self-consistency condition $\Delta_a=g\sum_m s_m^x$ reads
\beg
\sum_m \frac{e_m}{\sqrt{\eps_m^2+\Delta_a^2}}=\frac{2}{g}.
\label{gapeq}
\en
This is the BCS gap equation, which determines the value of $\Delta_a$ in the anomalous state.
The configuration of spins with all $e_m=1$   is equivalent to the BCS ground state. In this case $\Delta_a=\Delta_0$ -- the ground
state gap -- and \eref{gapeq} becomes in the continuum limit
\beg
\int_0^D \frac{d\eps}{\sqrt{\eps^2+\Delta_0^2}}=\frac{2}{\lam}\quad \lam=g \nu_F V\equiv \frac{g}{\delta},
\label{gapeqc}
\en
where $\nu_F$ is the density of states at the Fermi level and  $V$ is the volume of the system. In \eref{gapeqc} and
throughout this paper we assume the weak coupling regime $\Delta_0\ll D$ and
a constant density of $\eps_j$, $\nu(\eps)=\nu_F$, in the continuum limit. A non-constant density of states  modifies
the value of $\Delta_0$ determined from \eref{gapeqc} but will not affect any other equations derived
in the rest of the paper. As we will see,
these equations are confined to energies of order $\Delta_0$, while the density of states varies on an energy scale of order
$D$ or larger. Using $\Delta_0\ll D$, we obtain from \eref{gapeqc}
\beg
\Delta_0=2De^{-1/\lam}.
\label{gapvalue}
\en

The configuration with only one flipped spin, ${e_k=-1}$ and $e_{m\ne k}=1$,  corresponds to an excited state -- it contains  an excited
pair and has  energy
$2\sqrt{\eps_k^2+\Delta_0^2}$ relative to the ground state. Similarly, having two spins parallel to the field is equivalent to an
eigenstate with two excited pairs etc.

 Normal eigenstates are  spin arrangements where each spin is along $z$ axis, i.e.
 \beg\label{norm}
 2s_m^z=\pm1\equiv l_m,\quad s_m^-=0.
 \en
 They are also equilibria of the classical Hamiltonian \re{hcl} according to Eq.~\re{eqsm}. The Fermi ground
 state corresponds to $l_m=-\mbox{sgn } \eps_m$, while other normal eigenstates can be obtained from this state by moving  pairs of
 fermions from
 levels   below the Fermi energy to  levels   above it and flipping the corresponding spins.

\subsection{General properties of the dynamics}

In this subsection we introduce the Lax vector construction\cite{gensol1,dicke}, which plays a central role in analyzing the dynamics of
the BCS Hamiltonian.
We also define the separation variables and describe the general features of the dynamics.

The dynamics of the classical Hamiltonian \re{hcl} or, equivalently, Eqs.~\re{eqsm} and \re{bdg} turn out to be integrable. A convenient
tool for their analysis is the Lax vector defined as
\beg
{\bf L}(u)=- \frac{\hat {\bf z}}{g}+\sum_{m=1}^n \frac{{\bf s}_m}{u-\eps_m},
\label{lax}
\en
where $u$ is a complex parameter, $\hat {\bf z}$ is a unit vector along $z$ axis, and $n$ is the total
number of spins. The length of this vector is conserved by Eqs.~\re{eqsm} for any $u$, i.e.
\beg
\frac{d {\bf L}^2(u)}{dt}=0.
\label{cons}
\en
 For this reason ${\bf L}^2(u)$ can be viewed as the generator of the integrals of motion\cite{gensol1} for
Eqs.~\re{eqsm}. For example, its zeroes are conserved and constitute a set of independent integrals. Another possible choice for the
integrals
is e.g. the set of the residues of ${\bf L}^2(u)$ at the poles at $u=\eps_m$. Note that
\beg
{\bf L}^2(u)=\frac{ Q_{2n}(u)}{g^2\prod_j(u-\eps_j)^2},
\label{q}
\en
where $Q_{2n}(u)$ is a (spectral) polynomial of order $2n$. We also have
\beg
{\bf L}^2(u)=L_z^2(u)+L_-(u)L_+(u),
\label{l2}
\en
where $L_-(u)=L_x(u)-iL_y(u)$ and $L_{x,y,z}$ are the components of the Lax vector ${\bf L}(u)$.

To obtain solitons, we need to introduce new dynamical variables $u_m$\cite{sklyanin,vadim} in which Eqs.~\re{eqsm} separate and can be
integrated.
The separation variables are defined in terms of the ``old'' dynamical variables ${\bf s}_j$ as solutions of the following equation:
\beg
L_-(u_m) =\sum_{j=1}^n \frac{s_j^-}{u_m-\eps_j}=0.
\label{sep}
\en
This equation has $n-1$
solutions   since, when $L_-(u)$ is brought to a common denominator,  its numerator  is a polynomial of order $n-1$. Consequently,
there are $n-1$ separation variables $u_m$.
Eq.~\re{sep} can be inverted to obtain the spins in terms of the separation variables as follows
\beg
s_j^-=J_-\frac{\prod_{k}(\eps_j-u_k)}{\prod_{k\ne j}  (\eps_j-\eps_k)},
\label{su}
\en
where $J_-=J_x-i J_y$ as usual and ${\bf J}=\sum_j {\bf s}_j$ is the total classical spin.

In terms of the separation variables Eqs.~\re{eqsm} read
\beg\label{bcsev}
  \dot u_j=\frac{2i \sqrt{Q_{2n}(u_j)}}{\prod_{m\ne j}(u_j-u_m)},\quad j=1,\dots,n-1,
\en
\beg
  \dot J_-=-2i J_-\left(\sum_{j=1}^n  \eps_j+\frac{g J_z}{2}-\sum_{m=1}^{n-1}  u_m\right).
\label{jm}
\en

An important observation\cite{emilts} is that main properties of the dynamics   can be effectively discerned by analyzing
the zeros  of ${\bf L}^2(u)$. According to Eq.~\re{q}, these are the roots of the spectral polynomial $Q_{2n}(u)$ and we will often refer
to their configuration in the complex plane as to the {\it root diagram} of  ${\bf L}^2(u)$. Since $Q_{2n}(u)$ is positively defined, it
has $n$ pairs of complex conjugate roots.   For  generic initial conditions all $2n$ roots are distinct. In this case the dynamics
of the system is quasiperiodic with $n$  incommensurate frequencies and any dynamical quantity, e.g. the order parameter
$\Delta(t)=g J_-(t)$, typically contains all $n$ frequencies.

Significant simplifications occur when some  roots are  degenerate\cite{gensol,dicke}. It is important to distinguish between real and
complex
double roots. Note that any real root of $Q_{2n}(u)$ is automatically a double root (zero) because $Q_{2n}(u)$ is positively defined.
 A real  zero $c$ of  ${\bf L}^2(u)$ must also
be a zero of all three components of  ${\bf L}(u)$\cite{note1}. Further, note from Eq.~\re{sep} that one of the separation variables must
coincide with $c$.
In other words, it must be time-independent as it is ``frozen'' into the real root $c$. Eq.~\re{bcsev} shows that this is an
allowed solution of the equations of motion for the separation variables.
This freezing of a
separation variable can be translated into a genuine reduction of the number of degrees of freedom by one so that the dynamics of
the Hamiltonian \re{hcl} with $n$ spins  reduces to that of the same  Hamiltonian but with $n-1$ spins. In general,
$n-m$ real zeros (or equivalently $2m$ complex  zeros) mean a reduction of the dynamics  to that of $2m$ effective spins,
see Ref.~\onlinecite{gensol} and \onlinecite{dicke} for details.
Below
we will often encounter a situation when ${\bf L}^2(u)$ has a number of real zeros and consequently a number of separation variables
are frozen. The remaining variables we call unfrozen.

Let $u_{n-1}=c$ be the  separation variable frozen into the double zero of $Q_{2n}(u)$. Consider \eref{bcsev}
for $j\ne n-1$. Both the numerator
and the denominator of the right hand side contain a factor $u-c$, which  cancels lowering the order of the polynomial
under the square root by two. Suppose $Q_{2n}(u)$ has $n-2k$ double real zeros. Then, there are $2k-1$ unfrozen separation variables
$u_1,\dots,u_{2k-1}$. For these variables
 \eref{bcsev} can  be brought to the following form\cite{gensol} with the help of \eref{q}:
\beg
\sum_{j=1}^{2k-1} \frac{u_j^{l} du_j}{ \prod_m(u_j-\eps_m) \sqrt{ {\bf \widetilde{L} }^2(u_j) }}=2igdt\delta_{l,2k-2},
\quad l=0,\dots, 2k-2,
\label{sepeq}
\en
where ${\bf \widetilde{L} }^2(u)$ is obtained from ${\bf L}^2(u)$ by removing all real zeros $c_m$, i.e.
$$
{\bf \widetilde{L} }^2(u)= \frac{{\bf L}^2(u)}{\prod_m(u-c_m)^2}.
$$

\section{Linear analysis around   stationary states}
\label{linear}

In this section, we  analyze  equations of motion linearized  in the vicinity of normal and anomalous
stationary states. We show that the separation variables $u_j$ are the normal modes of the linearized problem. Some of the stationary
states are unstable. As we will see in the next section,
the corresponding normal modes become solitons in the nonlinear regime.

\subsection{Frequencies of oscillations around stationary states}
\label{sec:freq}

Here we show that the frequencies of small oscillations around normal  and anomalous states are determined by the zeros
of ${\bf L}^2(u)$, see also Ref.~\onlinecite{emilts}. When one of the frequencies is complex, the state is unstable and the corresponding
mode grows exponentially.

The linear analysis of equations of motion \re{eqsm} around stationary states greatly simplifies in terms of separation variables.
According to Eq.~\re{bcsev}, stationary  positions of $u_j$ are the roots of the polynomial $Q_{2n}(u)$ (or equivalently  the zeros
of  ${\bf L}^2(u)$,
see \eref{q}). Let us determine the form of  ${\bf L}^2(u)$ in the stationary states. Consider first the anomalous states \re{anom}.
Using Eqs.~\re{anom}, \re{lax}, and \re{gapeq}, we obtain
\beg
{\bf L}(u)=(\Delta_a \hat{\bf x}- u\hat{\bf z}) L_s(u),\quad {\bf L}^2(u)=(u^2+\Delta_a^2) L^2_s(u),
\label{lax0}
\en
where $\hat {\bf x}$ is a unit vector along $x$ axis and
\beg
L_s(u)=\sum_{m=1}^{n} \frac{e_m}{2(u-\eps_m)\sqrt{\eps_m^2+\Delta_a^2}},\quad e_m=\pm1.
\label{ls}
\en
Note that when the right hand side of Eq.~\re{ls} is brought to a common denominator,
the numerator is a polynomial of order $n-1$. Therefore, ${\bf L}^2(u)$ and consequently $Q_{2n}(u)$
have $n-1$ double zeros
$c_r$ -- the solutions of the equation $L_s(c_r)=0$. In addition, we see from Eq.~\re{lax0} that there are two roots
$u=\pm i\Delta_a$, i.e.
\beg\label{qfr}
Q_{2n}(u)=(u^2+\Delta_a^2)\prod_{r=1}^{n-1} (u-c_r)^2.
\en

When the spins ${\bf s}_j$ deviate from their equilibrium positions \re{anom}, the roots $c_r$ of polynomial $Q_{2n}(u)$ shift
to  $c_r+\delta c_r$\cite{note2}. Linearizing \eref{bcsev} in deviations $\delta c_r=a_r+ib_r$ and
$\delta u_r= u_r-c_r-a_r$ around the stationary positions $u_r=c_r$ and using \eref{qfr}, we obtain
\beg
\delta\dot u_r =2i \sqrt{c_r^2+\Delta_a^2}   \sqrt{(\delta u_r)^2 +b_r^2}
\en
with a solution $\delta u_r= b_r \sin[\w_r (t-t_0)]$, where $\w_r=2\sqrt{c_r^2+\Delta_a^2}$. Linearizing \eref{su}, one
 derives the spin variables in terms of $\delta u_r$. At this point we are   interested only in the frequencies $\w_r$.

We conclude that the separation variables are indeed the normal modes of the linearized problem (since they contain a single
frequency). The frequencies of small oscillations around anomalous stationary states are related to the double zeros $c_r$ of
${\bf L}^2(u)$ as $\w_r=2\sqrt{c_r^2+\Delta_a^2}$. If any of $\w_r$ has an imaginary part, the stationary state is unstable.

Next, consider linear analysis around normal eigenstates \re{norm}. In this case all spins are along $z$ axis. It follows
from Eqs.~\re{norm} and \re{lax} that ${\bf L}(u)=L_n(u)\hat {\bf z}$, where
\beg
L_n(u)=-\frac{1}{g}+\sum_{j=1}^n\frac{l_j}{2(u-\eps_j)},\quad l_j=\pm1.
\label{l0}
\en
We see that all zeros $c_r$ of ${\bf L}^2(u)=L_n^2(u)$ are double zeros. There are $n$ of them as the numerator of $L_n(u)$ is a
polynomial of order $n$, i.e. $Q_{2n}(u)= \prod_{r=1}^{n} (u-c_r)^2$. As before, the stationary positions of separation variables
  are $u_r=c_r$. Note however that there are only $n-1$ separation variables, so one of the $n$ zeros
$c_r$ must remain vacant.

Next, we show that the frequencies of small oscillations around a normal stationary state are $\w_r=2c_r$. When
one of the zeros $c_r$ is complex, the oscillatory behavior is replaced
with an  exponential growth, i.e. the stationary state is unstable.
Note that for small deviations from a normal state the $xy$ components of the total spin ${\bf J}$ are small.
Therefore,   in linear approximation  we can set the separation variables $u_j$ to their equilibrium values in Eqs.~\re{su}
and \re{jm},
$u_j=c_j$, i.e.   only $J_-$ is time-dependent.
As mentioned above,  $L_n(u)$ has a vacant zero (say $c_r$) which does not
correspond to any separation variable. \eref{jm} yields
\beg
 -\frac{d(\ln J_-)}{2i\, dt}=  \left[\sum_{j=1}^n  \eps_j+\frac{g J_z}{2} -\sum_{m=1}^{n}  c_m\right]+c_r=c_r,
\label{jmst}
\en
where we used the fact that the contribution in square brackets vanishes. This can be seen by observing that, since $c_m$ are the
zeros of $L_n(u)$, \eref{l0}
can be written as
\beg\label{lnst}
L_n(u)=-\frac{1}{g}\frac{\prod_m (u-c_m)}{\prod_j(u-\eps_j)}.
\en
Expanding the right hand sides of Eqs.~\re{lnst} and \re{l0} in $1/u$, matching the coefficients at $1/u$, and using $2J_z=\sum_j l_j$
(this follows from \eref{norm}), we see that the sum of terms in square brackets in \eref{jmst} is indeed zero. It follows
  that
$J_-\propto e^{-2ic_r t}$ and from \eref{su} we also derive $s_j^-\propto e^{-2ic_r t}$. Thus, the frequencies of oscillations
around normal stationary states are $\w_r=2c_r$.

\subsection{Examples of root diagrams of
 ${\bf L}^2(u)$
}
\label{rtex}

Here we provide examples of root diagrams  -- configurations of solutions of
the equation {${\bf L}^2(u)=0$} in the plane of complex $u$, see Ref.~\onlinecite{emilts} for more examples. We saw that
the zeros of ${\bf L}^2(u)$ evaluated
in stationary states determine the frequencies of oscillations around them. Moreover, the most important features of the dynamics, e.g.
the behavior of $\Delta(t)$ at large times, can be predicted by inspecting the root diagram\cite{emilts}, see also the discussion
below \eref{jm}. Similarly, we will see  that  the root diagram   determines the number and properties of
solitons corresponding to a given stationary state.

For simplicity, we assume particle-hole symmetry, i.e. the single fermion energies $\{\eps_m\}$ are symmetric with respect to
zero (Fermi level). According to \eref{anom},  this  means
\beg
\begin{split}
& s^x(\eps_m)=s^x(-\eps_m),\quad s^{z}(\eps_m)=-s^{z}(-\eps_m),\\
& s^{y}(\eps_m)=-s^{y}(-\eps_m),
\end{split}
\label{ph}
\en
where ${\bf s}_m\equiv {\bf s}(\eps_m)$. These relations can also be derived from \eref{bog} using particle-hole transformation
for fermion creation and annihilation operators  $\hat c_\sigma(-\eps_m) \leftrightarrow \hat c_\sigma^\dagger(\eps_m)$. Note
that relations \re{ph} are preserved by equations of motion \re{eqsm} and also imply $\Delta_y(t)=0$.

\begin{figure}[htb]
\includegraphics[width=0.65\textwidth]{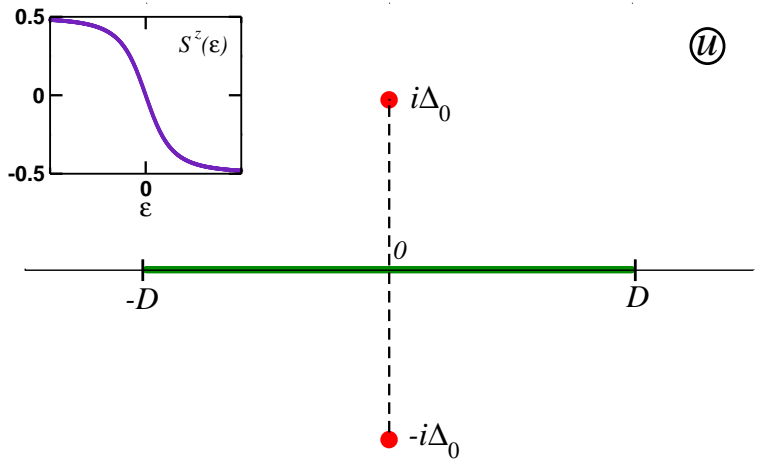}
\vspace{-2cm}
\caption{(color online) The roots  of ${\bf L}^2(u)=0$ (the root diagram) for the BCS ground state in the complex $u$ plane.
There is a line of double real roots (zeros of ${\bf L}^2(u)$) from $-D$ to $D$, where $D$ is the high-energy cutoff on
the single-fermion
states participating in the BCS Hamiltonian \re{bcs}. In addition, there are two imaginary single zeros $\pm i\Delta_0$,
where $\Delta_0$ is the ground state gap.
Frequencies $\w$ of small oscillations around the ground state are related to the zeros $c$ as $\w=\sqrt{c^2+\Delta_0^2}$,
see Sec.~\ref{bcs gr st}. We have $\w(\eps)=\sqrt{\eps^2+\Delta_0^2}$, where $-D\le \eps \le D$, and $\w=0$. The inset
shows the spin component $s^z(\eps)$ in the ground state. Since it has no discontinuities, there are no complex double zeros.%
}
\label{0afig}
\end{figure}

\subsubsection{ BCS ground state}
\label{bcs gr st}

 As discussed below \eref{gapeq}, the BCS ground state corresponds to $e_m=1$. We note from \eref{anom} that
spin components $s^x(\eps_m)$ and $s^z(\eps_m)$ in this state are continuous functions of
single particle energy $\eps_m$. It follows from Eqs.~\re{lax0} and \re{ls} that  ${\bf L}^2(u)=0$ has two single roots at
$u=\pm i\Delta_0$, see Fig.~\ref{0afig}, and $n-1$ double roots $c_k$ that are
the solutions of the following equation:
\beg\label{grroot}
L_s(u)=\sum_{m=1}^{n} \frac{1}{2(u-\eps_m)\sqrt{\eps_m^2+\Delta_0^2}}=0.
\en
All $n-1$ solutions are real. This can be seen by noting that  $L_s(u)$ changes
sign between consecutive $\eps_m$. Indeed, let $\eps_m$ be ordered so that $\eps_1<\eps_2<\dots<\eps_n$. Since $L_s(u)\to +\infty$
as $u\to\eps_m^+$ and $L_s(u)\to -\infty$ as $u\to\eps_{m+1}^-$, there is a point $u=c_m$ in the interval $(\eps_m, \eps_{m+1})$
where $L_s(c_m)=0$. According to the previous subsection, this yields a frequency $\w_m=2\sqrt{c_m^2+\Delta_0^2}$ of small oscillations
around the BCS ground state. In the continuum limit, when  level spacings $\eps_{m+1}-\eps_m$ tend to zero,
we have $c_m\approx\eps_m$ and  $\w_m\approx 2\sqrt{\eps_m^2+\Delta_0^2}$. In this limit, $c_m$ densely fill
the interval from $-D$ to $D=\max |\eps_m|$, i.e. ${\bf L}^2(u)$ has a line of double roots as shown
in Fig.~\ref{0afig}. Frequencies $\w_m$ are also the energies
of excited pairs  -- excitations
obtained by flipping the spin ${\bf s}_m$ from its ground state  position antiparallel to the
field ${\bf b}_m=(-2\Delta_0, 0, \eps_m)$ to
an equilibrium position parallel to the field ${\bf b}_m$, see Ref.~\onlinecite{ander1} for a discussion
of this relationship between the frequencies and the excitation spectrum.

\subsubsection{Excited anomalous states}
\label{ex an st}

 Next, consider two examples of anomalous stationary states  obtained from the
BCS ground state by flipping spins in  certain energy intervals.

\begin{figure}[htb]
\includegraphics[width=0.65\textwidth]{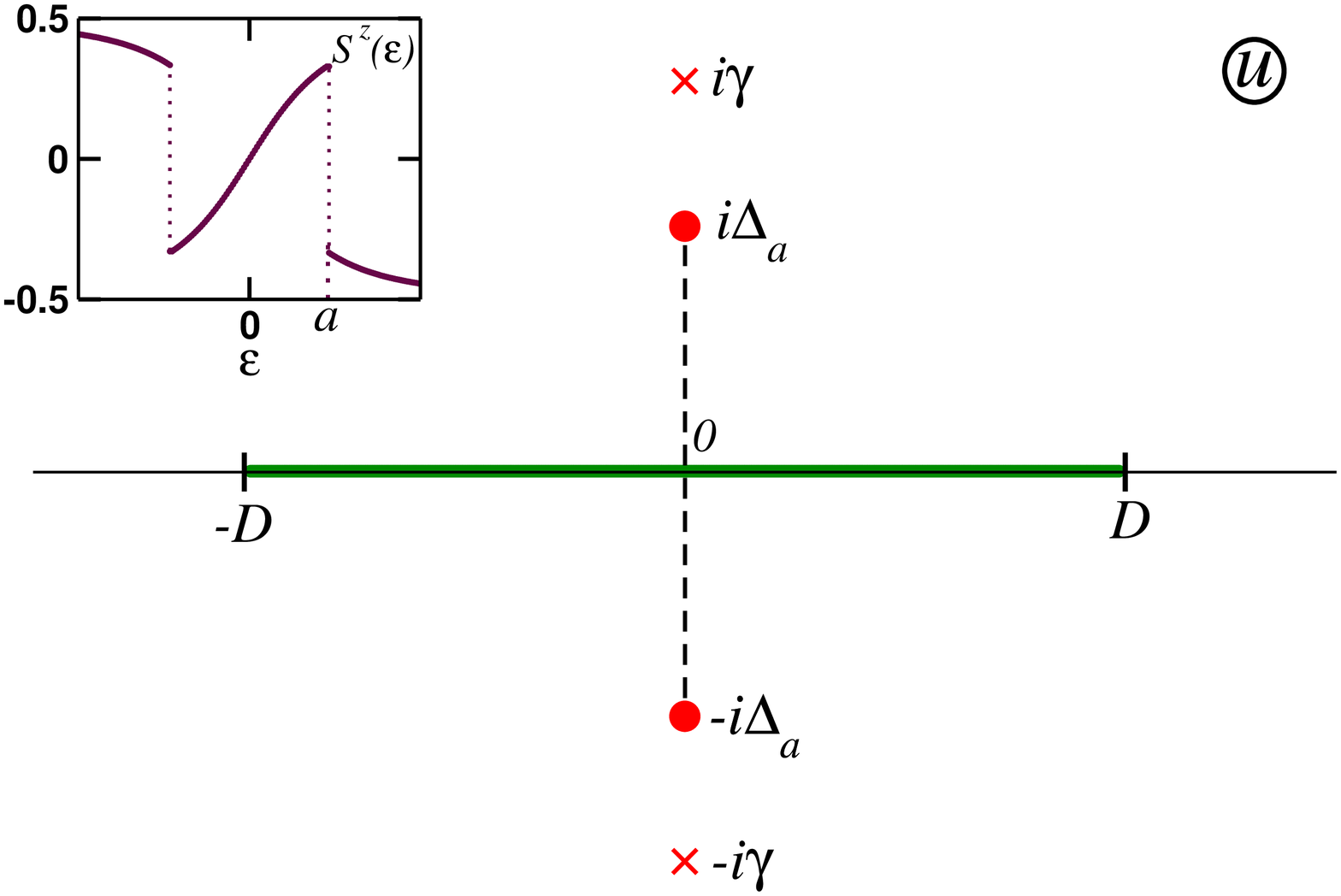}
\vspace{-1.3cm}
\caption{(color online) Zeros of  ${\bf L}^2(u)$  for an excited anomalous state in the complex $u$ plane. This anomalous
state has $2k=2$ jumps at $\eps=\pm a$ in the spin component $s^z(\eps)$ (inset) leading to $2k$ double imaginary zeros $\pm i\gamma$.
In addition, ${\bf L}^2(u)$ has
a line of real double zeros from $-D$ to $D$ and single zeros $\pm i\Delta_a$, where $\Delta_a$ is the value of the
order parameter in this state. In the present case, there is a single ($k=1$) unstable mode that grows with the rate
$\sqrt{\gamma^2-\Delta_a^2}$ giving rise to a single anomalous soliton shown in Fig.~\ref{1ansfig}, see Sec.~\ref{ex an st}.%
}
\label{1afig}
\end{figure}

\subsubsection*{Example 1}

 First, let spins in the interval $(-a,a)$ be flipped, i.e.
$e_m=\mbox{sgn}\, (|\eps_m|-a)$. This means that the Cooper pairs in this energy interval
are excited\cite{BCS,ander1}.   \eref{anom} implies that spin components $s^x(\eps)$ and $s^z(\eps)$ are
discontinuous  at $\eps=\pm a$ (see the inset in Fig~\ref{1afig}). As before   ${\bf L}^2(u)$ has two single zeros at $u=\pm i\Delta_a$
and $n-1$ double zeros
$c_k$ that are the solutions of $L_s(c_k)=0$, where $L_s(u)$ is given by \eref{ls}.
The difference is that in this case two of $c_k$ can be imaginary. Suppose $\eps_{m_1}<-a<\eps_{m_1+1}$ and $\eps_{m_2}<a<\eps_{m_2+1}$.
Then, $e_{m_1}=1$ while  $e_{m_1+1}=-1$ and similarly for $m_2$ and we are no longer guaranteed  real zeros of $L_s(u)$
in intervals $(\eps_{m_1},\eps_{m_1+1})$ and $(\eps_{m_2},\eps_{m_2+1})$ as in the BCS ground state. Instead, $L_s(u)$ can acquire
two complex conjugate zeros. In the particle-hole symmetric case $L_s(-u)=-L_s(u)$, which implies that these  zeros
must be purely imaginary as in Fig.~\ref{1afig}. The remaining $n-3$ double zeros of ${\bf L}^2(u)$ are real and lie between
consecutive $\eps_j$.
In the continuum limit, they merge into a continuous line of double zeros between $-D$ and $D$
as in the ground state.

To determine the two imaginary zeros $c=\pm i\gamma$ in the continuum limit, we rewrite the equation $L_s(u)=0$ in the form
\beg\label{1geq}
\int_0^\infty \frac{ \mbox{sgn} (\eps-a)\, d\eps}{(\eps^2+\gamma^2)\sqrt{\eps^2+\Delta_a^2}}=0,
\en
where we used $L_s(-u)=-L_s(u)$  and took the ultraviolet cutoff $D$ to infinity.
  In terms of
\beg\label{int}
F(\eps)=\int \frac{d\eps}{(\eps^2+\gamma^2)\sqrt{\eps^2+\Delta_a^2}}=\frac{1}{2\gamma\sqrt{\gamma^2-\Delta_a^2}}
\ln\left[
\frac{\gamma\sqrt{\eps^2+\Delta_a^2}+\eps\sqrt{\gamma^2-\Delta_a^2}}{\gamma\sqrt{\eps^2+\Delta_a^2}-
\eps\sqrt{\gamma^2-\Delta_a^2}}\right],
\en
 \eref{1geq} reads $F(+\infty)=2F(a)$. This equation has a unique positive solution
\beg
\label{1g}
\gamma=\left(a+\sqrt{a^2+\Delta_a^2}\right)\frac{a}{\Delta_a}.
\en

Note however that the gap equation \re{gapeq} has solutions only for sufficiently small  $a$.
To see this, we write down  \eref{gapeq} for the order parameter $\Delta_a$ in the anomalous state
where $e_m=\mbox{sgn} (|\eps_m|-a)$  and for the gap $\Delta_0$ in the BCS ground state where $e_m=1$.
Equating the left hand sides of the two equations, we obtain
\beg
\int_0^D \frac{d\eps}{\sqrt{\eps^2+\Delta_0^2}}=\int_0^D
\frac{ \mbox{sgn} (\eps-a)\, d\eps}{\sqrt{\eps^2+\Delta_a^2}}.
\label{2geq}
\en
In the
 $D\to\infty$ limit this equation yields
\beg
\label{dela}
\Delta_a^3-2\Delta_0\Delta_a^2+\Delta_0^2 \Delta_a-4 a^2=0
\en
together with the condition $\Delta_a<\Delta_0$.

The analysis of \eref{dela} shows that there are two solutions  $\Delta_a<\Delta_0$ provided $3\sqrt{3} a\le \Delta_0$ and no
solutions
  otherwise.  Interestingly, for one of the solutions $\gamma\le \Delta_a$, while for the other $\gamma\ge \Delta_a$. Which solution
  do we choose? Note that the quantum Hamiltonian \re{bcs} has $2^n$ unblocked states. Correspondingly,
  there are $2^n$ choices of $e_m=\pm1$. More than one solution for a given selection of $e_m$ means
  that we have more states in the mean-field than there are eigenstates
of the original quantum Hamiltonian. It is natural to expect that among the two solutions for $\Delta_a$ the one
that yields a stable anomalous state corresponds to the quantum eigenstate.
We have shown above that frequencies of small oscillations around anomalous stationary states
are related to the zeros $c_k$  as $\w_k=\sqrt{c_k^2+\Delta_a^2}$. For the zeros $\pm i\gamma$ we have
$\w_\gamma=i\sqrt{\gamma^2-\Delta_a^2}$. We see that for $\gamma>\Delta_a$ the frequency is imaginary and the corresponding normal
mode grows exponentially. Therefore,  the solution $\Delta_a<\gamma$ yields an unstable
anomalous state, while for $\Delta_a>\gamma$ we get a stable state. Both states however can play an important role
in the description of the dynamical problem \re{bdg}.

\begin{figure}[htb]
\includegraphics[width=0.65\textwidth]{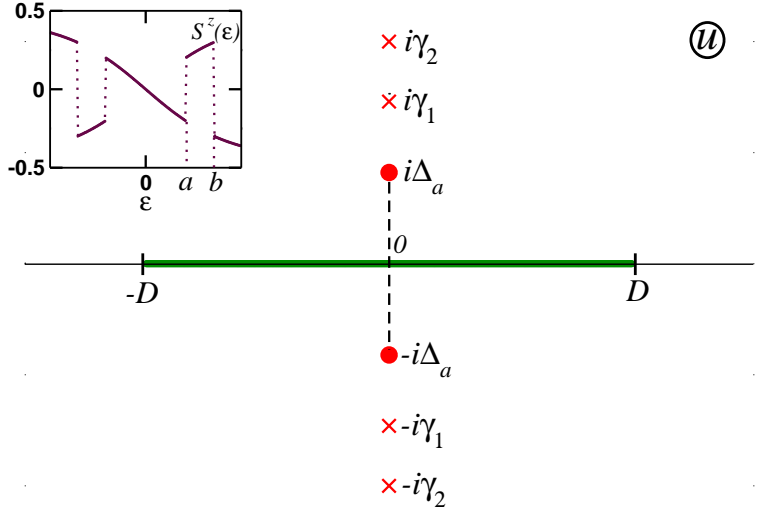}
\vspace{-1.3cm}
\caption{(color online) Zeros of  ${\bf L}^2(u)$  for an excited anomalous state in the complex $u$ plane. This anomalous
state has $2k=4$ discontinuities  at $\eps=\pm a$ and $\pm b$ in the spin component $s^z(\eps)$ (inset) leading to $2k$ double imaginary
zeros $\pm i\gamma_{1,2}$.
In addition, ${\bf L}^2(u)$ has
a line of real double zeros from $-D$ to $D$ and single zeros $\pm i\Delta_a$, where $\Delta_a$ is the value of the
order parameter in this state. In the present case, there are $k=2$ unstable modes that grow with rates
$\sqrt{\gamma_{1,2}^2-\Delta_a^2}$ giving rise to 2-anomalous-solitons shown in Fig.~\ref{2asfig}, see Sec.~\ref{ex an st}.%
}
\label{4afig}
\end{figure}

\subsubsection*{Example 2}

 A more involved example of an anomalous state is obtained by flipping spins in two energy intervals, e.g.
in intervals $(-b,-a)$ and $(a,b)$ symmetric with respect to the Fermi level. This implies
$e_m=\mbox{sgn} (|\eps_m|-a)(|\eps_m|-b)$. Now the spin components have four discontinuities at $\eps=\pm a$
and $\eps=\pm b$ (inset in  Fig.~\ref{4afig}). Correspondingly, ${\bf L}^2(u)$ can have four complex double zeros $\pm i\gamma_{1,2}$
as in Fig.~\ref{4afig} in addition to two single
zeros at $u=\pm i\Delta_a$ and $n-5$ real zeros on the line from $-D$ to $D$. This follows in a manner similar
to the above analysis of the state with $e_m=\mbox{sgn} (|\eps_m|-a)$. In general, $2k$ discontinuities in spin components in an
anomalous stationary state
can lead  to $2k$ complex double roots.

To determine the complex zeros  in the continuum limit, we repeat the procedure
that lead to \eref{1g}. Now we derive $2F(b)-2F(a)=F(+\infty)$, where $F(\eps)$ is given by expression \re{int}. This equation has
solutions $\pm i\gamma_{1,2}$, where
\beg
\label{2g}
4\frac{\gamma_{1,2}}{\Delta_a}=xy-1\pm\sqrt{ (x y+1)^2-4(x+y-1)},
\en
 $x\Delta_a^2=\left(\sqrt{a^2+\Delta_a^2}-a\right)^2$, and $y\Delta_a^2=\left(\sqrt{b^2+\Delta_a^2}+b\right)^2$.
The gap equation \re{gapeq} in terms of $x$ and $y$ takes the form
\beg
xy=\frac{\Delta_0}{\Delta_a},
\label{56}
\en
where $\Delta_0$ is the
ground state gap. Using these equations, it is not difficult to select $a$ and $b$ so that $\gamma_{1,2}$ are
real and $\gamma_2>\gamma_1>\Delta_a$ as shown in Fig.~\ref{4afig}.
This is the choice we will need in Sec.~\ref{2anomsec}.

\begin{figure}[htb]
\includegraphics[width=0.65\textwidth]{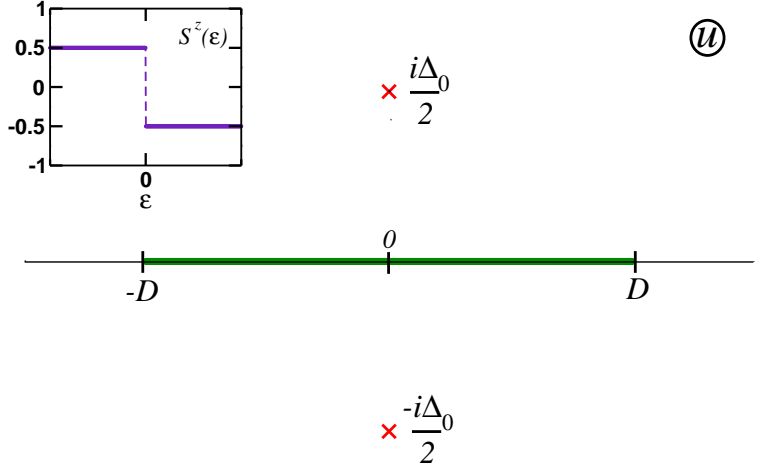}
\vspace{-2cm}
\caption{(color online) Zeros of ${\bf L}^2(u)$ for the Fermi ground state in the complex $u$ plane. The single
discontinuity ($k=1$)
in the spin component $s^z(\eps)$ at $\eps=0$, see the inset, leads to a single ($2k-1=1$) pair  of complex double zeros at
$\pm i\Delta_0/2$,
where $\Delta_0$ is the order parameter in the BCS ground state. There is also a line of double real zeros stretching from
$-D$ to $D$, where $D$ is the high-energy cutoff on the single-fermion
states participating in the BCS Hamiltonian \re{bcs}. Frequencies  $\w$ of small oscillations around this state
are related to the zeros $c$ as $\w=2c$, see Sec.~\ref{sec:freq}. In the present case, there is a single ($k=1$) unstable mode
that grows with the rate $\gamma=\Delta_0$ giving rise to $k=1$ normal-soliton shown in Fig.~\ref{1nsfig}%
}
\label{1nfig}
\end{figure}

\subsubsection{Fermi ground state}
\label{fermi gr st}

 We saw  that in normal stationary
states all zeros of ${\bf L}^2(u)$ are double degenerate and are solutions
of the equation $L_n(u)=0$, see \eref{l0}. The Fermi ground state
has all states below the Fermi energy occupied and  states
above it empty. This corresponds to
$2s_j^z=l_j=-\mbox{sgn}\,\eps_j$. Therefore, the zeros are determined by
the following equation:
\beg
\sum_{j=1}^n\frac{\mbox{sgn}\,\eps_j}{u-\eps_j}=-\frac{2}{g}.
\label{fgs}
\en
 There are $n$ solutions each one being a double
zero of ${\bf L}^2(u)$. The  analysis of \eref{fgs} is similar to that of \eref{grroot}.
\eref{fgs} has real roots between consecutive $\eps_j$  except
when $\mbox{sgn}\,\eps_j$ changes from 1 to $-1$. Therefore, there is a real root $c_j$ in each interval $(\eps_j, \eps_{j+1})$ except
for the interval  containing the Fermi level. Since there are
$n-2$ such intervals, $n-2$ roots are real while the remaining two can be complex. Due to the particle-hole symmetry \re{ph}
the complex roots must be purely imaginary, see Fig.~\ref{1nfig}. They also must be complex conjugate to each other as \eref{fgs} is
invariant under
complex conjugation.

In the continuum limit the spacing between $\eps_j$ vanishes and for the real roots
we have $c_j\approx \eps_j$, i.e. ${\bf L}^2(u)$ has a line of double real zeros stretching from $-D$ to $D$,
Fig.~\ref{1nfig}.
To determine the two imaginary roots $\pm i\gamma$, we rewrite \eref{fgs} in the integral form
$$
\int_0^D \left( \frac{d\eps}{\eps-i\gamma}+  \frac{d\eps}{\eps+i\gamma}\right)=\frac{2}{\lam}.
$$
Using \eref{gapvalue}, we obtain in the weak coupling regime $\Delta_0\ll D$
\beg
\gamma=\frac{\Delta_0}{2}.
\label{fgsrt}
\en
Thus, according to the discussion in the previous subsection, equations of motion \re{eqsm} linearized around the Fermi ground state
show $n-1$ stable modes with oscillation frequencies $\w_j\approx 2\eps_j$ and one unstable mode that grows as
$e^{2\gamma t}=e^{\Delta_0 t}$\cite{Elihu,single}. Note also that
 the $z$ component of spins $2s^z(\eps_j)=-\mbox{sgn}\,\eps_j$
in the Fermi ground state experiences a single jump at the Fermi level.

\begin{figure}[htb]
{\bf a)}\includegraphics[width=0.45\textwidth]{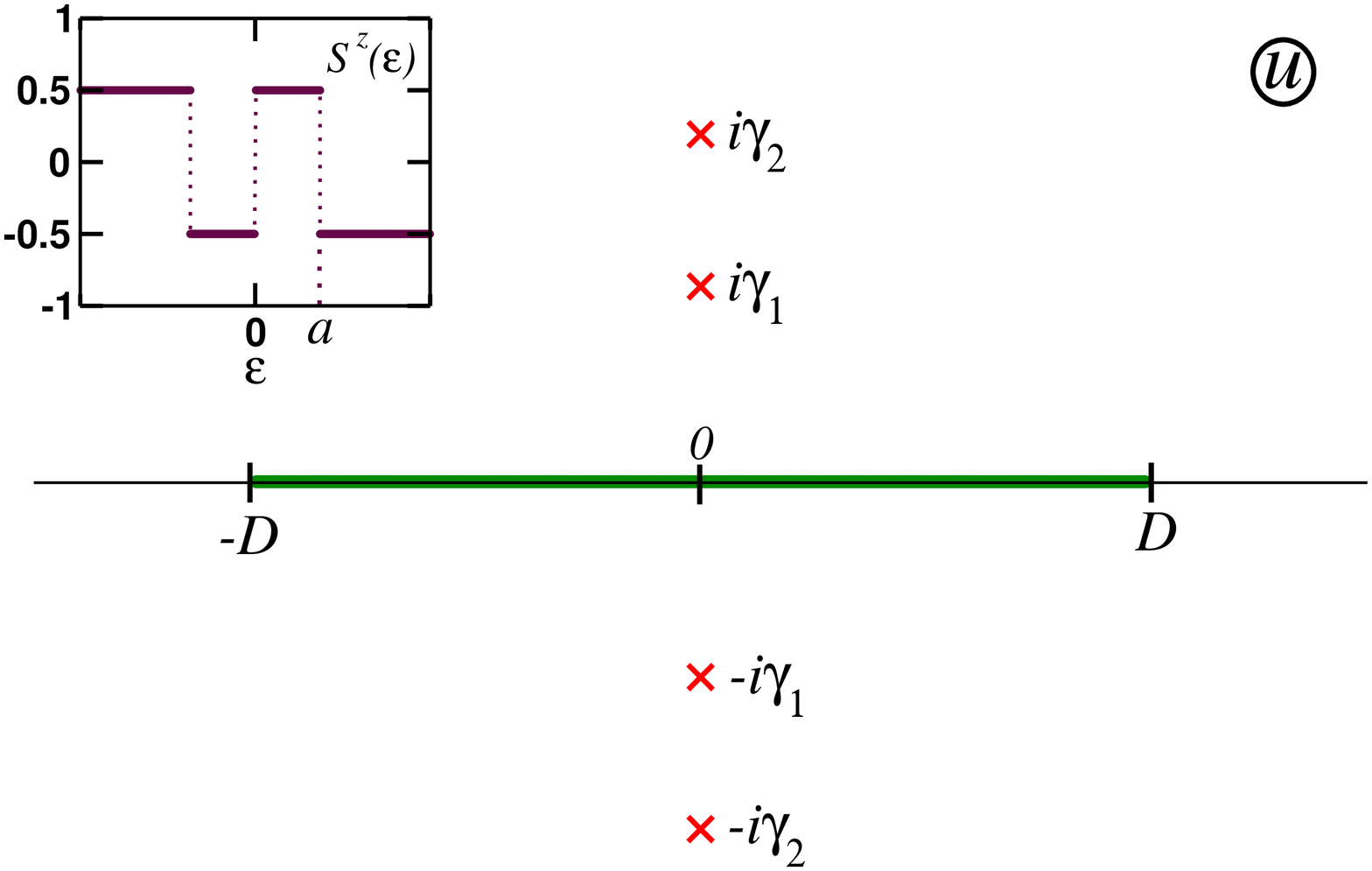}$\qquad$
{\bf b)}\includegraphics[width=0.45\textwidth]{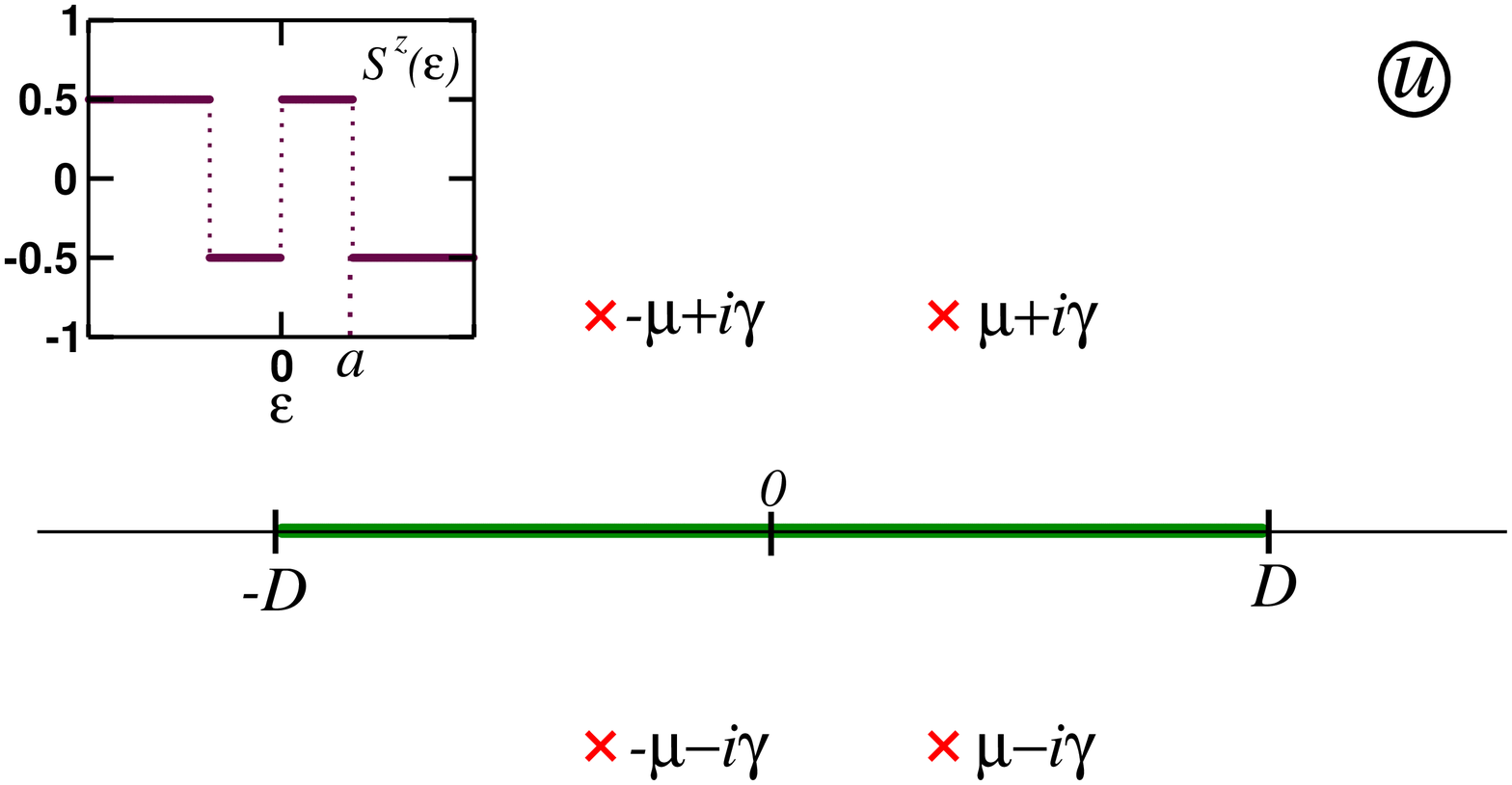}
\caption{(color online) Zeros of ${\bf L}^2(u)$ for an excited normal state characterized by $2k-1=3$ jumps
in the spin component $s^z(\eps)$ at $\eps=\pm a$ (insets). There are $k=2$ pairs of complex conjugate
double zeros (identified with crosses)
leading
to $k=2$ normal-soliton solutions shown in Fig.~\ref{2nsfig}. The complex zeros are determined by $a$ and the ground state gap
$\Delta_0$, see \eref{44}. a) corresponds to the case $4a<\Delta_0$ and  Fig.~\ref{2nsfig}a) and b) corresponds to $4a>\Delta_0$ and
Fig.~\ref{2nsfig}b).%
}
\label{3nfig}
\end{figure}

\subsubsection{Excited normal state}
\label{ex n st}

 Now consider  a normal stationary state where $s^z(\eps_j)$ has three discontinuities.
We require $2s^z(\eps_j)=-1$ (1) for large positive (negative) $\eps_j$. Otherwise, the first term
in \eref{hcl} is not minimized at large $\eps_j$ and single particle states far from the Fermi level are affected by the pairing
interaction, which is unphysical. Under these conditions the total number of discontinuities in $s^z(\eps_j)$ must be odd.
Therefore, the next option after the Fermi ground state that has one jump is a state with three jumps in
$s^z(\eps_j)$.

Let $2s_j^z=l_j=-\mbox{sgn}\,\eps_j(\eps_j^2-a^2)$, i.e. spins in the interval $|\eps_j|\le a$
point in  directions opposite to those  in the Fermi ground state, see the insets in Fig.~\ref{3nfig}. The solutions
of ${\bf L}^2(u)=0$ are determined
in the same way as for the Fermi ground state. In the present case, we find that there are $n-4$ double roots located
between  $\eps_j$ and $\eps_{j+1}$ except when $l_j$ and $l_{j+1}$ have different signs. The remaining four double roots
can take complex values.
In the continuum limit, ${\bf L}^2(u)$ has a line of double zeros from $-D$ to $D$ and four isolated complex zeros
$c=i\gamma_{1,2}$ and $c=-i\gamma_{1,2}$ shown in Fig.~\ref{3nfig}, where
\beg
\gamma_{1,2}=\frac{\Delta_0}{4}\pm \sqrt{\frac{\Delta_0^2}{16}-a^2}.
\label{44}
\en
Correspondingly, there are two unstable modes (one for each pair of complex conjugate  zeros). If $a\le \Delta_0/4$,
\eref{44} yields real $\gamma_{1,2}$ (Fig.~\ref{3nfig}a) and the unstable modes grow as $e^{2\gamma_1t}$ and $e^{2\gamma_2t}$.
For $a>\Delta_0/4$ we have $c=\pm\mu\pm i\gamma$ (Fig.~\ref{3nfig}b), where $\mu=\Delta_0/4$ and
$\gamma=\sqrt{a^2-\Delta_0^2/16}$. In
this case  unstable  modes diverge in an oscillatory manner as $e^{\pm 2i\mu t} e^{2\gamma t}$. In general, a normal stationary
state with $2k-1$ discontinuities in $s^z(\eps_j)$ is characterized by up to $2k$ complex double zeros of ${\bf L}^2(u)$ and $k$
unstable modes.

\section{Normal solitons}
\label{normal}

In this section, we determine solutions of equations of motion \re{bdg}  that asymptote to normal stationary states
at $t\to\pm\infty$. In particular, we derive equations \re{1ns} -- \re{sumns} for normal solitons, see also
Figs.~\ref{1nsfig}, \ref{2nsfig}, and \ref{3nsfig}. That these solutions are
solitons can be seen in a number of ways. First, these are trajectories that connect an unstable equilibrium to itself, i.e. they
start in an unstable stationary state at $t\to -\infty$ and return to it at $t\to\infty$. This is typical of solitons\cite{Arnold},
see the
paragraph preceding \eref{1ns}. Second, as we will show, in a certain regime the solution splits into a sum of single solitons
as it should\cite{Fadeev}.
   Finally, in contrast to the general solution these solutions are in terms
of elementary functions\cite{Novikov}.

\subsection{General $k$-normal-soliton solution}
\label{gennormal}

Consider a general normal stationary state with $2k-1$ discontinuities in $s^z(\eps_j)$. Suppose ${\bf L}^2(u)=L_n^2(u)$ has $2k$ complex
double zeros $c_1,c_2,\dots, c_{2k}$, i.e. there are $k$ unstable modes in the linear analysis.
Let us solve equations of motion for separation variables \re{sepeq} for this state.

First, we derive a useful equation for the time-dependent gap function $\Delta(t)$.
  $L_n(u)$ has $2k$ complex conjugate zeros $c_1,c_2,\dots, c_{2k}$
  and $n-2k$ real zeros.
Bringing \eref{l0} to a common denominator, we obtain
\beg
L_n(u)=-\frac{1}{g}\frac{P_{2k}(u) R_{n-2k}(u)}{\prod_m (u-\eps_m)},
\label{lnpoly}
\en
where $P_{2k}(u)=\prod_{r=1}^{2k}(u-c_r)$ and $R_{n-2k}(u)$ represents the contribution of the real zeros. Both these
polynomials have real coefficients.
\eref{l2} yields
\beg
\left[ L_z(u)-L_n(u)\right] \left[ L_z(u)+L_n(u)\right]=-L_-(u) L_+(u)
\label{decomp}
\en
This equation implies
\beg
\begin{array}{l}
\dis L_z(u)+L_n(u)=-\frac{2}{g} \frac{ S_k(u) S^*_k(u)R_{n-2k}(u)}{\prod_m (u-\eps_m)},\\
\\
\dis L_z(u)-L_n(u)=-\frac{g}{2}J_-J_+ \frac{ T_{k-1}(u) T^*_{k-1}(u)R_{n-2k}(u)}{\prod_m (u-\eps_m)},\\
\end{array}
\label{lzpmln}
\en
\beg
L_-(u)= J_- \frac{S_k(u) T_{k-1}(u)R_{n-2k}(u)}{\prod_m (u-\eps_m)},
\label{l-poly}
\en
where $S_k(u)$ and $T_{k-1}(u)$ are polynomials in $u$ of orders $k$ and $k-1$, respectively. The coefficient at highest
power of $u$ is equal to unity in all polynomials. The coefficients of $S^*_k(u)$ are complex conjugate to those of $S_k(u)$ and
similarly for $T^*_{k-1}(u)$. The prefactors in Eqs.~\re{lzpmln} and \re{l-poly} are obtained from large $u$ behavior.
For example, Eqs.~\re{lax} and \re{l0} imply $L_z(u)+L_n(u)\approx -2/g$ at $u\to\infty$. The right hand side of
the first equation in \re{lzpmln} has the same large $u$ asymptote. The polynomial $R_{n-2k}(u)$ is common to all
components of ${\bf L}(u)$ since any real zero of ${\bf L}^2(u)$ is also a zero of $L_{x,y,z}$ \cite{note1}.
Subtracting the second equation in \re{lzpmln} from the first one and using \eref{lnpoly}, we derive
\beg
P_{2k}(u)=S_k(u)S^*_k(u)+\frac{|\Delta|^2}{4} T_{k-1}(u) T^*_{k-1}(u),
\label{real}
\en
where we used $\Delta(t)=g J_-(t)$. We will need this equation below to determine $|\Delta(t)|$.

Now consider equations of motion \re{sepeq}. It follows from \eref{l2} and the definition \re{sep}
of separation variables $u_j$ that ${\bf L}^2(u_j)=L_z^2(u_j)$. Comparing the large $u$ behavior of $\sqrt{{\bf L}^2(u)}\to
1/g$
and $L_z(u)\to -1/g$, we conclude that $\sqrt{{\bf L}^2(u_j)}=-L_z(u_j)$.
\eref{sepeq} takes the following form:
\beg
\begin{array}{l}
\dis\sum_{j=1}^{2k-1} \frac{u_j^{l} du_j}{ \widetilde{L}_z(u_j) \prod_m (u_j-\eps_m) }=0,\quad l=0,\dots, 2k-3,\\
\\
\dis \sum_{j=1}^{2k-1} \frac{u_j^{2k-2} du_j}{ \widetilde{L}_z(u_j) \prod_m (u_j-\eps_m) }=-2i g dt,\\
\end{array}
\label{jip}
\en
where $\widetilde{L}_z(u)=L_z(u)/R_{n-2k}$.  Eqs.~\re{sep} and \re{l-poly} imply that unfrozen separation
variables $u_1,\dots,u_{2k-1}$ are the roots of $S_k(u)T_{k-1}(u)$. Let $u_1,\dots,u_{k-1}$ be the roots of $T_{k-1}(u)$
and $u_k,\dots,u_{2k-1}$ the roots of $S_k(u)$. \eref{lzpmln} reads $\widetilde{L}_z(u_j)=\widetilde{L}_n(u_j)$ for
$j=1,\dots,k-1$ and $\widetilde{L}_z(u_j)=-\widetilde{L}_n(u_j)$ for $j=k,\dots,2k-1$, where $\widetilde{L}_n(u)=L_n(u)/R_{n-2k}$.
Further, using \eref{lnpoly}, we obtain from \eref{jip}
\beg
\sum_{j=k}^{2k-1} \frac{u_j^{l} du_j}{ P_{2k}(u_j)}-\sum_{j=1}^{k-1} \frac{u_j^{l} du_j}{ P_{2k}(u_j)}=-2i dt \delta_{l, 2k-2},
\quad l=0,\dots,2k-2.
\label{nsin1}
\en

\eref{nsin1} does not contain square roots in contrast to \eref{sepeq} and can be integrated in elementary functions.
To do so, we expand the ratios $u^l/P_{2k}(u)$ in elementary fractions
\beg
\frac{u^l}{ P_{2k}(u)}=\sum_{m=1}^{2k} \frac{c_m^l}{(u-c_m) \prod_{j\ne m } (c_m-c_j)},\quad l<2k.
\label{ident}
\en
This identity can be verified by comparing residues at poles  $u=c_m$ on both sides. Using expansion \re{ident}
in \eref{nsin1}, we obtain
\beg
\sum_{m=1}^{2k} \frac{c_m^l dx_m}{\prod_{j\ne m}(c_m-c_j)}=-2i dt \delta_{l, 2k-2},
\label{nsin2}
\en
where $l=0,\dots,2k-2$ and
\beg
dx_m=\sum_{j=k}^{2k-1} \frac{du_j}{u_j-c_m}-\sum_{j=1}^{k-1}\frac{du_j}{u_j-c_m}=d\ln  \frac{S_k(c_m)}{T_{k-1}(c_m)},\quad
x_m\equiv \frac{S_k(c_m)}{T_{k-1}(c_m)}.
\label{nsin3}
\en
Integration of \eref{nsin2} results in
\beg
\sum_{m=1}^{2k} \frac{c_m^l x_m}{\prod_{j\ne m}(c_m-c_j)}=-2i t \delta_{l, 2k-2}+E(c_l), \quad l=0,\dots,2k-2,
\label{nsin4}
\en
where $E(c_l)$ are the integration constants. These equations are linear in $x_m$ with the general
solution
\beg
x_m=-2ic_m t+\widetilde{E}(c_m)+G(t).
\label{xm}
\en
$\widetilde{E}(c_m)$ are new time-independent constants and $G(t)$ is an arbitrary function of $t$.

Using the definition of $x_m$ in \eref{nsin3}, we find
\beg
\frac{S_k(c_m)}{T_{k-1}(c_m)}=-A(c_m) F(t) e^{-2ic_m t}, \quad m=1,\dots,2k,
\label{exm}
\en
where $A(c_m)$ are complex constants and $F(t)$ is a function of time to be determined below. Eqs.~\re{exm} are
$2k$ linear equations for $2k-1$ coefficients of polynomials $S_k(u)$ and $T_{k-1}(u)$. The compatibility
condition for this linear system yields a linear equation for the function $F(t)$.
We derive
\beg
F(t)=(-1)^k2^{2k-2}  \frac{D_{k-1}}{D_k},
\label{F}
\en
where the determinant $D_r$ is given by
\beg
D_r=
\left|
\begin{array}{lll}
f & \dots & f^{(r-1)}\\
\vdots & & \vdots\\
f^{(r-1)}&\dots & f^{2(r-1)}\\
\end{array}
\right|,
\label{gapnormaltxt}
\en
$f^{(j)}$ is the $j$th derivative of the function $f(t)$ with respect to $t$, and
\beg
f(t)=\sum_{m=1}^{2k} \frac{ A(c_m) e^{-2i c_m t} }{\prod_{l\ne m}(c_m-c_l)}.
\label{ftxt}
\en

To relate $|F(t)|$ to $|\Delta(t)|$, we use \eref{real}. This equation also imposes certain restrictions
on complex constants $A(c_m)$. Setting $u=c_m$ in \eref{real}  and using the fact that $c_m$ are the roots
of $P_{2k}(u)$, $P_{2k}(c_m)=0$, we obtain
\beg
\frac{S(c_m)}{T(c_m)} \frac{S^*(c_m)}{T^*(c_m)}=-\frac{|\Delta|^2}{4}.
\label{deltast}
\en
Note that while the coefficients of the polynomial $S_m^*(u)$ are
complex conjugate to those of $S_m(u)$, $S^*(c_m)$ is not complex conjugate to $S(c_m)$ since $c_m$ is complex.
Instead, we have $S^*(c_m)=[S(c_m^*)]^*$, i.e. $S^*(c_m)$ is conjugate to $S(c_m^*)$. Using this and \eref{exm},
we obtain from \eref{deltast}
\beg
A(c_m) A^*(c_m^*) |F(t)|^2 =-\frac{|\Delta(t)|^2}{4},
\label{aa}
\en
where $A^*(c_m^*)$ is the complex conjugate of $A(c_m^*)$ -- the constant corresponding to the zero
$c_m^*$ complex conjugate to $c_m$ (recall that the zeros $c_j$ of ${\bf L}^2(u)$ come in complex conjugate pairs).
\eref{aa} implies that the product $A(c_m) A^*(c_m^*)$ is independent of $m$. With no loss of generality we set
\beg
A(c_m) A^*(c_m^*)=-1.
\label{aa1}
\en
Any other real value will rescale $|F(t)|$ without affecting $|\Delta(t)|$. Therefore, we have $|\Delta(t)|=2|F(t)|$ and
\beg
|\Delta(t)|=2^{2k-1}\left| \frac{D_{k-1}}{D_k}\right|.
\label{deltext}
\en
It follows from \eref{aa1} that the constants $A(c_m)$ can be parameterized as follows
\beg
A(c_l)=e^{\alpha_l+i\phi_l},\quad A(c_{k+l})=-e^{-\alpha_l+i\phi_l},
\label{Atxt}
\en
where $\alpha_l$ and $\phi_l$ are arbitrary real parameters and we ordered the $2k$ zeros $c_m$
 so that $c_{k+l}=c_l^*$ and $\mbox{Im}(c_l)>0$ for
$l=1,\dots,k$.

Eqs.~\re{deltext}, \re{gapnormaltxt}, \re{ftxt}, \re{exm} and \re{Atxt} fully describe the general $k$-normal-soliton solution
(examples for $k=1,2$, and 3 are shown in Figs.~\ref{1nsfig}, \ref{2nsfig}, and \ref{3nsfig}, respectively).
They contain
$2k$ zeros $c_m$ fixed by the normal stationary state corresponding to this solution. This state has $2k-1$ discontinuities
in $s^z(\eps_m)$. The zeros $c_m$ are the roots of the equation $L_n(u)=0$, where $L_n(u)$ is given by \eref{l0}.
  The $2k$ real
parameters $\alpha_l$ and $\phi_l$ in \eref{Atxt} are arbitrary. That the general $k$-normal-soliton should be
indeed characterized by $2k$ arbitrary real parameters is seen from the discussion in the paragraph following \eref{jm}.
As mentioned there (see  Refs.~\onlinecite{gensol} and \onlinecite{dicke} for details), a real double zero of ${\bf L}^2(u)$
effectively reduces the number of degrees of freedom (spins) by one.
Since in the present case we have $n-2k$ such roots, it can be described by $2k$ effective spins. Then, there are $4k$ initial
conditions
(two angles per each spin).  $2k$ of these are determined by the $2k$ integrals of motion $c_m$, while the other $2k$ correspond
to $\alpha_l$ and $\phi_l$.

\subsection{Matching soliton constants to spin configuration at large negative time}

Here we show that the  $k$-soliton \re{gapnormaltxt} tends to
a normal stationary state in $t\to\pm\infty$ limits  and relate the constants $\alpha_l$ and $\phi_l$ to the
deviations of spins from this state at large negative times.

First, let us evaluate  expression \re{gapnormaltxt} for large negative $t$. To this end, we keep in \eref{ftxt}
only the exponents that diverge in the $t\to-\infty$ limit, i.e. the $k$ terms that have $\mbox{Im}(c_m)<0$.
After some manipulations with the rows of determinants $D_k$ and $D_{k-1}$, we derive
\beg
|\Delta(t)|=2\biggr|\sum_{m=1}^k
\frac{ e^{-2ic_mt} e^{\alpha_m+i\phi_m} (c_m-c^*_m)\prod_i (c_m-c^*_i) }{\prod_{i\ne m} (c_m-c_i)}\biggl|.
\label{gaplarget}
\en

We see that $|\Delta(t)|\propto e^{-2\gamma t}$ at large negative $t$, where $\gamma$ is
the minimum of $|\mbox{Im}(c_m)|$. Quantities $J_\pm(t)$ and $F(t)$ behave in the same way as they are
proportional to $|\Delta(t)|$.   According to \eref{exm} as $t\to -\infty$ either $S_k(c_m)\to 0$
or $T_{k-1}(c_m)\to 0$ except for the zero $c_m$ with $\mbox{Im}(c_m)=i\gamma$.
Since  the unfrozen separation variables are the roots of either $S_k(u)$ or $T_{k-1}(u)$ (see Eqs.~\re{sep} and \re{l-poly}), they
must tend to their stationary state positions $c_m$. Observe also that
since $J_\pm(t)\to 0$, the second equation in \re{decomp} and \eref{l-poly} mean $L_z(u)\to L_n(u)$ and $L_-(u)\to0$.
It follows from the definition \re{lax} of ${\bf L}(u)$ and \eref{l0} that $s_j^{x,y}\to 0$ and $s_j^z\to l_j/2$. Thus,
the $k$-normal-soliton tends to the normal stationary state that has the same values of zeros $c_i$. The analysis
of the $t\to\infty$ limit is completely analogous.

Next, consider the limiting  stationary state.  There are $2k$ zeros $c_i$ and only $2k-1$ unfrozen separation variables,
i.e. one of the zeros $c_i$ (say $c_r$) remains vacant.
Suppose spins deviate from this stationary state keeping the values of
integrals of motion $c_i$  the same.
Since $J_-=0$ in normal states, \eref{su} yields to the linear order in the deviation
\beg
s_j^-=J_-\frac{R_{n-2k}(\eps_j)\prod_{i}(\eps_j-u_i^{(0)})}{\prod_{i\ne j}(\eps_j-\eps_i)},
\label{sult}
\en
where $u_i^{(0)}=c_i$ are the stationary positions of the unfrozen separation variables and
$R_{n-2k}(\eps_j)$ is the contribution of the frozen ones. The frozen variables are located in  real zeros of
$L_n(u)$, which are also the zeros of $R_{n-2k}(u)$, see \eref{lnpoly}.
Further,  Eqs.~\re{l0} and \re{lnpoly}
imply
\beg
-\frac{1}{g}+\sum_{j=1}^n\frac{l_j}{2(u-\eps_j)}=-\frac{1}{g}\frac{ R_{n-2k}(u)(u-c_r) \prod_i(u-u_i^{(0)}) }{\prod_m (u-\eps_m)}.
\label{lnln}
\en
Equating the residues at poles  $u=\eps_j$ on both sides, we obtain
$$
\frac{R_{n-2k}(\eps_j)\prod_{i}(\eps_j-u_i^{(0)})}{\prod_{i\ne j}(\eps_j-\eps_i)}=-\frac{l_j g}{2(\eps_j-c_r)}.
$$
Substituting this into \eref{sult}, we find
\beg
s_j^-(t)=-g J_-(t)\frac{l_j}{2(\eps_j-c_r)}.
\label{slin}
\en

Finally, $J_-(t)$ is determined from \eref{jmst}, which was also derived in a linear analysis around normal
stationary states. The difference is that there we considered generic deviations  when
the integrals of motion $c_i$ also deviate from their stationary state values. Nevertheless, \eref{jmst} is the
same in both cases and integrating it, we obtain
\begin{align}
\label{ltmode}
& \Delta(t)=g J_-(t)=\beta_r e^{-2i c_r t},\\
\label{sjmode}
& s_j^-(t)=- \beta_r  \frac{l_j e^{-2i c_r t}}{2(\eps_j-c_r)}.
\end{align}
These are particular solutions of the linearized equations of motion. They describe an unstable mode with
complex ``frequency'' $2c_r$.

The general solution (with $c_i$ fixed to their stationary state values) is a superposition of all modes,
i.e.
\begin{align}
\label{ltgen}
& \Delta(t)=g J_-(t)=\sum_{r=1}^k\beta_r e^{-2i c_r t},\\
\label{sjgen}
& s_j^-(t)=- \sum_{r=1}^k\beta_r  \frac{l_j e^{-2i c_r t}}{2(\eps_j-c_r)}.
\end{align}
Note that these equations contain only $c_r$ such that $\mbox{Im}(c_r)<0$, same as in \eref{gaplarget}, to insure that
the deviations are indeed small at large negative $t$. Comparing Eqs.~\re{ltgen} and \re{gaplarget}, we find
\beg
\beta_m=\frac{2  (c_m-c^*_m)\prod_i (c_m-c^*_i) }{\prod_{i\ne m} (c_m-c_i)}e^{\alpha_m+i\phi_m}, \quad \mbox{Im}(c_m)<0.
\label{ba}
\en
Eqs.~\re{ba} and \re{sjgen} specify deviations of spins from their normal stationary state positions necessary to
generate the $k$-normal-soliton \re{deltext}. Indeed, an arbitrary choice of real $\alpha_m$, $\phi_m$, and large
negative $t=t_0$
determines $\beta_m$ and  deviations of spins \re{sjgen}. Equations of motion \re{eqsm} started at $t=t_0$
with these initial conditions produce the $k$-soliton solution \re{deltext} with the same
values of $c_i$ as those in the stationary state. On the other hand, note that generic deviations
of spins will modify $c_i$, see e.g. the text following \eref{qfr}, and will not lead to  solitons.

\subsection{Examples of 1 and 2-normal-solitons}

In this subsection, we consider $k=1$ and $k=2$ normal solitons in more detail, see also the Introduction.

{\it 1-normal-soliton.} The single normal soliton solution (Fig.~\ref{1nsfig}) has been previously found in Ref.~\onlinecite{single}.
Here
we derive it from the general $k$-soliton \re{deltext} as its simplest particular case to illustrate our construction of
multi-soliton solutions.
In this case $k=1$ and ${\bf L}^2(u)$ has two complex double zeros $c_1=c_2^*\equiv \mu+i\gamma$ as illustrated
in Fig.~\ref{1nfig}.
The corresponding normal stationary state has a single discontinuity in the $z$ component of spin (inset
in Fig.~\ref{1nfig}), i.e. it is
the Fermi ground state, see Sec.~\ref{rtex}. We have seen that in the particle-hole symmetric case
$2s_j^z=-\mbox{sgn}\eps_j$, $\mu=0$, and $\gamma=\Delta_0/2$.

Eqs.~\re{aa}, \re{ftxt}, and \re{F} yield
$$
A(c_1)=e^{\alpha+i\phi},\quad A(c_2)=-e^{-\alpha+i\phi},\quad F=-i\frac{\gamma}{\cosh(2\gamma t+\alpha)}e^{2i\mu t -i\phi}.
$$
and
\beg
|\Delta(t)|=2|F(t)|=\frac{2\gamma}{\cosh(2\gamma t+\alpha)}.
\label{nsing}
\en
Graphically, the single soliton  is represented by a single peak located at $t_0=-\alpha/2\gamma$,
see Fig.~\ref{1nsfig}.
The parameter $\gamma$ controls the width and the height of the peak.

There is $2k-1=1$ unfrozen separation variable $u_1$. Therefore,
$S_k(u)=u-u_1$ and $T_{k-1}(u)=1$. \eref{exm} implies
\beg
u_1(t)=\mu-i\gamma \tanh(2\gamma t+\alpha).
\label{u1}
\en
Note that $u_1\to \mu\pm i\gamma =c_{1,2}$ as $t\to\mp\infty$ in agreement with the results of the previous subsection.
The separation variable starts from  the complex zero $\mu+i\gamma$ of ${\bf L}^2(u)$ at $t=-\infty$ and goes to
the complex conjugate zero $\mu-i\gamma$ at $t=\infty$ along the straight line connecting the two zeros shown
in Fig.~\ref{1nfig}.

Individual spin components can be derived from Eqs.~\re{lnpoly}, \re{l-poly}, and \re{u1}. We have
$$
L_n(u)=-\frac{1}{g}\frac{(u-c_1)(u-c^*_1) R_{n-2}(u)}{\prod_m(u-\eps_m)},\quad
L_-(u)=J_-\frac{(u-u_1)  R_{n-2}(u)}{\prod_m(u-\eps_m)}.
$$
Therefore,
$$
L_-(u)=-\Delta(t)\frac{(u-u_1)}{(u-c_1)(u-c_1^*)}L_n(u).
$$
Using expression \re{l0} with $l_j=-\mbox{sgn}\eps_j$ and comparing the residues
at poles at $u=\eps_j$ on both sides of the above equation, we obtain\cite{single}
$$
s_j^-(t)=s_j^x(t)+is_j^y(t)=\Delta(t)\frac{\left[\eps_j-u_1(t)\right]\mbox{sgn }\eps_j }{2\left[(\eps_j-\mu)^2+\gamma^2\right]}.
$$
Similarly, the second equation in \eref{lzpmln} yields
$$
s_j^z(t)=\frac{\mbox{sgn }\eps_j}{2}\left[ \frac{|\Delta(t)|^2}{(\eps_j-\mu)^2+\gamma^2}-1\right].
$$

{\it 2-normal-soliton.} Now $k=2$ and  ${\bf L}^2(u)$ has  four complex zeros, Fig.~\ref{3nfig}. The limiting excited
normal state exhibits $2k-1=3$ jumps in $s^z(\eps_j)$, see the inset in Fig.~\ref{3nfig}. In Sec.~\ref{rtex}, we considered  such a
stationary state with $2s_j^z=-\mbox{sgn}\,\eps_j(\eps_j^2-a^2)$ and determined the corresponding complex zeros.

For $a\le \Delta_0/4$ these zeros are purely imaginary (Fig.~\ref{3nfig}a), $c_1=i\gamma_1$, $c_2=i\gamma_2$, $c_3=-i\gamma_1$,
and $c_4=-i\gamma_2$, where $\gamma_{1,2}$ are given by \eref{44}.  \eref{deltext} yields
\beg
|\Delta(t)|=A\left| \frac{h(t)}{h(t)\ddot h(t) -\dot h^2(t)}\right|, \label{2nstxt}
\en
where $A=4|\gamma_2^2-\gamma_1^2|$ and
\beg
 h(t)=e^{i\phi_1}\frac{\cosh(2\gamma_1t+\alpha_1)}{2\gamma_1}+e^{i\phi_2}\frac{\cosh(2\gamma_2t+\alpha_2)}{2\gamma_2}.
\label{h1txtx}
\en
The plot of the 2-normal-soliton \re{2nstxt} displays two peaks, see Fig.~\ref{2nsfig}. Parameters $\alpha_{1,2}$ determine the
location of the
peaks in time, while $\gamma_{1,2}$ control their widths and heights. The 2-soliton can be viewed as a nonlinear
superposition of two single solitons. At large  separation between solitons in time, $|\alpha_1-\alpha_2|\gg 1$,
we obtain from \eref{2nstxt}
\beg
|\Delta(t)| \approx\frac{2\gamma_1}{\cosh(2\gamma_1 t+\alpha_1+\eta)}+\frac{2\gamma_2}{\cosh(2\gamma_2 t+\alpha_2-\eta)},
\label{2nslargesep}
\en
where the phase shift $\eta$ is
$$
\tanh\eta=\mbox{sgn}(\alpha_2-\alpha_1)\frac{2\gamma_1\gamma_2}{\gamma_1^2+\gamma_2^2}.
$$
In deriving \eref{2nslargesep} we neglected the terms of relative smallness $e^{-|\alpha_1-\alpha_2|}$. We see that
at large separation, the 2-normal-soliton reduces to a simple sum of two single solitons as shown
in Fig.~\ref{2nsfig}. This is a general
property of solitons and one can show that the general $k$-normal-soliton \re{deltext} also obeys this rule,
see e.g. Fig.~\ref{3nsfig} and \eref{sumns}. For small separation the two peaks merge into one.

When $a>\Delta_0/4$ in \eref{44} the four roots of ${\bf L}^2(u)$ have the form $\pm \mu\pm i\gamma$ (Fig.~\ref{3nfig}b), where
$\mu=\Delta_0/4$ and $\gamma=\sqrt{a^2-\Delta_0^2/16}$.
 In this case the 2-normal soliton is again given by \eref{2nstxt} where now $A=16\mu\sqrt{\mu^2+\gamma^2}$ and
\beg
h(t)=e^{-2i\mu t+i\phi_1}\frac{\cosh (2\gamma t+\alpha_1-i\beta)}{2\gamma}+
e^{2i\mu t+i\phi_2}\frac{\cosh (2\gamma t+\alpha_2+i\beta)}{2\gamma}.
\label{h2}
\en
An additional feature as compared to \eref{h1txtx} is that here the two terms ``rotate'' with frequency
$4\mu$ with respect to one another. For large separation, $|\alpha_1-\alpha_2|\gg 1$, this has no effect --
the plot of $|\Delta(t)|$ still shows two peaks well separated in time, dashed lines in Fig.~\ref{2nsfig}b. Now the peaks are the same, i.e.
$\gamma_1=\gamma_2=\gamma$ in \eref{h1txtx}. In contrast, when the separation is small there is a single peak
as in the 2-soliton \re{h1txtx} but
with an amplitude  modulated by an oscillation with  frequency $\w\sim 4\mu=\Delta_0$, see Fig~\ref{2nsfig}b.

\section{Anomalous solitons}
\label{anomalous}

In this section, we construct 1- and 2-anomalous-solitons (see also Figs.~\ref{1ansfig} and \ref{2asfig}) -- solutions of
Bogoliubov-de Gennes equations for $|\Delta(t)|$ that
asymptote to anomalous stationary states \re{anom} as $t\to\pm\infty$. These solutions show the same solitonic signatures as
normal solitons, see the introductory paragraph in Sec.~\ref{normal}.
In particular, they are expressed in terms of exponentials and
multi-solitons break up into a  sum of well separated single anomalous solitons in a certain limit.

\subsection{Single anomalous soliton as a special case of a 3-spin solution}

A single soliton corresponds to the anomalous state with  one unstable mode in the linear analysis, i.e. ${\bf L}^2(u)$
has two double complex zeros in addition to single zeros $u=\pm i\Delta_a$, see Sec.~\ref{sec:freq} and
Fig.~\ref{1afig}.
We considered a state of this type in Sec.~\ref{rtex}. In this example, spins in the energy interval $(-a,a)$ are flipped;
$e_m=\mbox{sgn} (|\eps_m|-a)$ in \eref{anom} as shown in Fig.~\ref{1afig} (inset). In other words, Cooper pairs for single particle
states $-a\le \eps\le a$ are excited.
This state is particle-hole symmetric \re{ph} and the complex zeros of ${\bf L}^2(u)$ are therefore purely imaginary,
$u=\pm i\gamma$. As we have shown in Sec.~\ref{sec:freq}, this anomalous state is unstable for $\gamma>\Delta_a$.

For the particle-hole symmetric case equations of motion \re{eqsm} have the following form:
\beg
\dot s_j^x=-2\eps_j s_j^y,\quad \dot s_j^z=-2\Delta s_j^y,\quad \dot s_j^y=2\Delta s_j^z+2\eps_j s_j^x,
\label{phem}
\en
where $\Delta=g\sum_j s_j^x$ is real since $\sum_j s_j^y=0$ at all times. Let us solve \eref{phem} under the condition
that at $t\to -\infty$ the solution asymptotes to the above anomalous state. As mentioned below \eref{sep} and
detailed in Refs.~\onlinecite{gensol} and \onlinecite{dicke}, when ${\bf L}^2(u)$ has $m$ complex conjugate zeros
(the remaining $2n-2m$ zeros are real) the problem is reduced to solving equations of motion \re{phem} for
$m$ effective spins. In the present case $m=3$ (counting the pair of double zeros as two pairs) and therefore we will need to solve
\eref{phem} for three spins.

This reduction can be seen in Eqs.~\re{bcsev} and \re{jm}. Suppose $Q_{2n}(u)$ has only three pairs of complex conjugate
roots $(c_1, c^*_1), (c_2, c^*_2)$, and $(c_3, c_3^*)$.
There are only $3-1=2$ unfrozen separation variables, while
the remaining $n-3$ are frozen into the $n-3$ double real roots of $Q_{2n}(u)$, see the text
following \eref{sep}. Suppose $c$ is  a real root and let $u_{n-1}=c$.
Then $Q_{2n}(u_j)$ contains a factor $(u_j-c)^2$ which cancels $u_j-u_{n-1}=u_j-c$ in the denominator of \eref{bcsev}.
This cancellation occurs for all frozen separation variables and we obtain
\begin{align}
\label{bcsev3}
&  \dot u_j=\frac{2i \sqrt{Q_{6}(u_j)}}{\prod_{m\ne j}(u_j-u_m)},\quad j,m=1,2,\\
\label{jm3}
&  \dot J_-=2i J_-( u_1+u_2),
\end{align}
where $Q_6(u)=\prod_{i=1}^3(u-c_i)(u-c_i^*)$. \eref{jm3} follows from \eref{jm}, since $\sum_j\eps_j$, $J_z$, and the sum of
frozen separation variables, $\sum_{j=3}^{n-1} u_j$   vanish due to the particle-hole symmetry\cite{note3}. We see that
equations of motion \re{bcsev3} and \re{jm3} are exactly the same as \re{bcsev} and \re{jm} for $n=3$ in the particle-hole
symmetric case. Since the latter equations and \eref{phem} are equivalent, Eqs.~\re{bcsev3} and \re{jm3}
describe the motion of three effective spins ${\bf S}_1$, ${\bf S}_2$, and ${\bf S}_3$. Note that $J_-(t)$ and
consequently $\Delta(t)=gJ_-(t)$ are the same in both problems. Moreover, one can show\cite{gensol,dicke} that
the original spins are linearly related to the effective ones, i.e.
\beg
{\bf s}_j=a_j {\bf S}_1+b_j {\bf S}_2+d_j {\bf S}_3
\label{3spin}
\en
Thus, to construct a single anomalous soliton, we need to solve \eref{phem} for three spins.

First, let us obtain a general
3-spin solution for which $Q_6(u)$ has three distinct pairs of  complex conjugate roots. As discussed above, the soliton
corresponds to the special case when two of these pairs, $\pm i\gamma$, are degenerate. The third pair is $u=\pm i\Delta_a$
and therefore $Q_6(u)=(u^2+\gamma^2)^2(u^2+\Delta_a^2)$. The particle-hole symmetry of the 3-spin problem implies
$\eps_1=-\eps$, $\eps_2=0$, $\eps_3=\eps$ and
\beg
S_1^x=S_3^x\equiv S_x,\quad - S_1^{y,z}=S_3^{y,z}\equiv S_{y,z},\quad S_2^x=-\frac{1}{2},\quad S_2^{y,z}=0.
\label{ph3}
\en
Using $\Delta=g\sum_{m=1}^3 S_m^x=2g S_x-g/2$ and integrating \eref{phem}, we determine
the effective spins
\beg
S_x=\frac{\Delta}{2g}+\frac{1}{2},\quad S_y=-\frac{\dot\Delta}{4g\eps},\quad S_z=\frac{\Delta^2}{4g\eps}+ C,
\label{3spdelt}
\en
where $C$ is an integration constant. Combining Eqs.~\re{3spdelt} and \re{3spin}, we derive the original spins
in terms of $\Delta(t)$,
\beg
s_j^x=A_j\Delta+F_j, \quad s_j^y=B_j\dot \Delta,\quad s_j^z=C_j\Delta^2+D_j,
\label{3spor}
\en
where $A_j$, $B_j$, $C_j$, $D_j$, and $F_j$ are time-independent. The constants $B_j$, $C_j$, and $D_j$ are odd in $\eps_j$,
while $A_j$ and $F_j$ are even by particle-hole symmetry \re{ph}, i.e. $B_j\equiv B(\eps_j)=-B(-\eps_j)$ etc.
Since $\Delta=g\sum_j s_j^x$ we also  have
\beg
g\sum_{j=1}^n A_j=1,\quad  \sum_{j=1}^n F_j=0.
\label{constr}
\en
\eref{3spor} is similar to the ansatz of Ref.~\onlinecite{single}, which is obtained by setting $F_j=0$. Nevertheless,
this difference is important as this ansatz yields  2-spin solutions\cite{gensol}, while here we construct 3-spin ones.

Substituting \eref{3spor} into equations of motion \re{phem}, we find
\beg
\begin{array}{l}
\dis A_j=-2\eps_j B_j,\quad C_j=-B_j,\quad F_j=\frac{2c_1 B_j}{\eps_j},\quad D_j=2(\eps_j^2-c_2)B_j\\
\\
\dis B_j=-\frac{\eps_j e_j}{4\sqrt{Q_6(\eps_j)}}, \quad Q_6(u)= u^2(u^2-c_2)^2+c_1^2-c_3 u^2,\\
\end{array}
\label{abcdf}
\en
where $e_j=\pm1$. Since $B_j$ is odd, $e_j$ must be even,  $e(\eps_j)=e(-\eps_j)$. The polynomial $Q_6(u)$ is
the same spectral polynomial that appears in \eref{bcsev3}, see the discussion of few spin solutions
in Refs.~\onlinecite{gensol} and \onlinecite{dicke}.
The coefficients of the polynomial $Q_6(u)$ are constrained by \eref{constr}. Plugging \eref{abcdf} into
\eref{constr}, we obtain
\beg
\sum_{j=1}^n \frac{e_j}{\sqrt{Q_6(\eps_j)}}=0,\quad \sum_{j=1}^n \frac{\eps_j^2 e_j}{\sqrt{Q_6(\eps_j)}}=\frac{2}{g},
\label{const1}
\en
which provides two constraints on three parameters $c_1$, $c_2$, and $c_3$. Thus, 3-spin solutions constructed here
are a one parameter family of solutions to \eref{phem}.

It remains to determine $\Delta(t)$ for 3-spin solutions. The equation for $\Delta(t)$ can be obtained from
the condition that the length of spins is conserved by the evolution, ${\bf s}_j^2=1/4$. With the help of Eqs.~\re{3spor}
and \re{abcdf} this condition reduces to
\beg
\dot\Delta^2=-P_4(\Delta),\quad P_4(\Delta)=\Delta^4+4c_2\Delta^2-8c_1\Delta+4c_3.
\label{delta3sp}
\en
For general $P_4(\Delta)$ the solution of this equation is an elliptic function. Here we are only interested in an anomalous
soliton. As discussed above, it corresponds to a special choice of the spectral polynomial
$Q_6(u)=(u^2+\gamma^2)^2(u^2+\Delta_a^2)$. According to the expression for $Q_6(u)$ in \eref{abcdf}, this implies
$$
c_1=-\gamma^2\Delta_a,\quad c_2=-\frac{\Delta_a^2}{2}-\gamma^2,\quad c_3=\frac{\Delta_a^4}{4}-\gamma^2\Delta_a^2.
$$
For these values of the parameters, \eref{delta3sp} for the order parameter takes the form
\beg
\dot\Delta^2=-(\Delta-\Delta_a)^2(\Delta^2+2\Delta\Delta_a+\Delta_a^2-4\gamma^2).
\label{3spsol}
\en
Now the fourth order polynomial on the right hand side has a double root $\Delta_a$, which means that \eref{3spsol}
can be solved by elementary means. Note also that the stationary state value $\Delta(t)=\Delta_a$ is also a solution.
In terms of a new variable $y=(\Delta-\Delta_a)^{-1}$ \eref{3spsol} reads $\dot y^2=\lam^2y^2-4\Delta_a y-1$, where
$\lam=2\sqrt{\gamma^2-\Delta_a^2}$. We obtain
\beg
\Delta(t)-\Delta_a=\frac{ \lam^2}{2\Delta_a\pm 2\gamma \cosh (\lam t+\alpha)}, \quad \lam=2\sqrt{\gamma^2-\Delta_a^2}.
\label{deltaanomtxt}
\en
The constraints \re{const1} become the gap equation \re{gapeq} and the equation determining imaginary
zeros  $\pm i\gamma$ of ${\bf L}^2(u)$
$$
\sum_j \frac{e_j}{(\eps_j^2+\gamma^2)\sqrt{\eps_j^2+\Delta_a^2}}=0,\quad   \sum_j \frac{e_j}{\sqrt{\eps_j^2+\Delta_a^2}}=\frac{2}{g}.
$$
We solved these equations in Sec.~\ref{rtex}, see Eqs.~\re{1g} and \re{dela}.

Finally, \eref{3spor} together with Eqs.~\re{abcdf} and \re{deltaanomtxt} yield the  individual spin components for
a single anomalous soliton
\beg
\begin{array}{l}
\dis s_j^x(t)=\frac{e_j\eps_j^2(\Delta(t)-\Delta_a)}{2(\eps_j^2+\gamma^2)\sqrt{\eps_j^2+\Delta_a^2}}+
\frac{e_j\Delta_a}{2\sqrt{\eps_j^2+\Delta_a^2}},\\
\\
\dis s_j^y(t)=-\frac{e_j\eps_j \dot\Delta(t)}{4(\eps_j^2+\gamma^2)\sqrt{\eps_j^2+\Delta_a^2}},\\
\\
\dis s_j^z(t)=\frac{e_j\eps_j(\Delta^2(t)-\Delta_a^2)}{4(\eps_j^2+\gamma^2)\sqrt{\eps_j^2+\Delta_a^2}}-
\frac{e_j\eps_j}{2\sqrt{\eps_j^2+\Delta_a^2}}.\\
\end{array}
\label{indvsp}
\en
Eqs.~\re{deltaanomtxt} and \re{indvsp}   describe a single anomalous soliton solution to the equations
of motion \re{bdg} and \re{eqsm}. Note that for $t\to\pm\infty$ the order parameter $\Delta(t)\to\Delta_a$
and the spin components tend to their
stationary state values \re{anom}. For $\Delta_a=0$ the anomalous soliton \re{deltaanomtxt} turns into the normal
one \re{nsing}. Graphically, the anomalous soliton  is represented by a single peak
similarly to the normal soliton, see Fig.~\ref{1ansfig}.
Parameters
$\Delta_a$ and $\gamma$ control its width and height, while $\alpha$ determines its position in time.


\subsection{2-anomalous-soliton solutions}
\label{2anomsec}

Two and higher anomalous solitons can be derived by solving equations of motion for the separation variables \re{sepeq}
similarly to the construction of normal solitons above. The $k$-anomalous-soliton corresponds to a root diagram of
${\bf L}^2(u)$ with $k$ double complex zeros $c_1,\dots,c_k$ and a pair of single zeros $\pm i\Delta_a$. Then,
 the denominator of \eref{sepeq} is $\sqrt{u^2+\Delta_a^2}\prod_i(u-c_i)$, i.e. only a second order polynomial
 remains under the square root. In this case, \eref{sepeq} can be integrated in elementary functions. However, here
 we restrict ourselves to the 2-soliton case and adopt a simpler approach to construct it.

Spin
components of the anomalous stationary state, to which the $k$-soliton asymptotes at large times, display
$2k$ discontinuities.  We analyzed an example with four discontinuities in Sec.~\ref{rtex}, see also
Fig.~\ref{4afig}. In this example
spins in energy intervals $(-b,-a)$ and $(a,b)$ are flipped, which means $e_m=\mbox{sgn} (|\eps_m|-a)(|\eps_m|-b)$
in \eref{anom}.
${\bf L}^2(u)$ has four  complex double zeros $\pm i\gamma_{1,2}$ in addition to single zeros $\pm i\Delta_a$
as illustrated in Fig.~\ref{4afig}.
We assume $\gamma_{2}>\gamma_1>\Delta_a$. Therefore, there are two unstable modes in linear analysis with growth rates
$\lam_{1,2}= 2(\gamma^2_{1,2}-\Delta_a^2)^{1/2}$, see Sec.~\ref{sec:freq}. The 2-anomalous soliton must have the
following properties: a) $\Delta(t)\to\Delta_a$ as $t\to\pm\infty$ while at large $t$ it should reproduce the linear
analysis, b) for $\Delta_a=0$ it should be equivalent to the 2-normal-soliton described by Eqs.~\re{2nstxt} and \re{h1txtx},
and c) in a certain regime the 2-soliton must break up into a sum of two single solitons \re{deltaanomtxt}. This suggests
the following ansatz for the 2-soliton
\begin{align}
\label{delans}
&\Delta(t)-\Delta_a=\frac{f}{f\ddot f-\dot f^2},\\
 &f=a_0
+\frac{a_1}{\lam_1}\cosh(\lam_1t+\alpha_1)+\frac{a_2}{\lam_2}\cosh(\lam_2t+\alpha_2),
\label{fansatz}
\end{align}
where $a_0$, $a_1$, and $a_2$ are time-independent parameters. To determine them, we require that for
$|\alpha_2-\alpha_1|\gg 1$ the 2-soliton  be well approximated by a sum of two single anomalous solitons (cf. \eref{2nslargesep})
\beg
\Delta(t)-\Delta_a\approx\frac{\lam_1^2}{2\Delta_a\pm 2\gamma_1 \cosh (\lam_1 t+\alpha_1+\eta)}+
\frac{\lam_2^2}{2\Delta_a\pm 2\gamma_2 \cosh (\lam_2 t+\alpha_2-\eta)}
\label{want}
\en
Neglecting terms of relative smallness $e^{-|\alpha_1-\alpha_2|}$ in \eref{delans}, we indeed obtain \eref{want}
when $\tanh (\eta/2)=\lam_1/\lam_2$ and
\beg
f=\frac{2\Delta_a}{\lam_1^2\lam_2^2}\pm \frac{2\gamma_1}{\lam_1^2(\lam_2^2-\lam_1^2)}\cosh(\lam_1t+\alpha_1)
\pm \frac{2\gamma_2}{\lam_2^2(\lam_2^2-\lam_2^2)}\cosh(\lam_2t+\alpha_2), \quad \lam_{1,2}= 2\sqrt{\gamma^2_{1,2}-\Delta_a^2}.
\label{f2an}
\en
Further, one can verify that the 2-anomalous-soliton given by Eqs.~\re{delans} and \re{f2an} also has the properties
a) and b) discussed above. Its plot consists of two peaks levelling off to the stationary value $\Delta_a$
at large times, see Fig.~\ref{2asfig}. The amplitudes and the widths of these peaks are determined by parameters
$\gamma_1$, $\gamma_2$, and $\Delta_a$.

\section{Conclusion}

In this paper, we constructed soliton solutions of time-dependent Bogoliubov-de Gennes equations \re{bdg} or, equivalently,
Gorkov equations \re{eqsm} describing the collisionless dynamics of a fermionic superfluid. There are two types of solitons.
Normal solitons asymptote at $t\to\pm\infty$ to  normal stationary states
 \re{norm}, which are  simultaneous eigenstates of the mean-field BCS Hamiltonian \re{bcs1} and the Fermi gas. These states
 are characterized by zero order parameter, $\Delta=0$. We have derived the general $k$-normal-soliton solution, Eqs.~\re{gapnormaltxt},
 \re{ftxt}, and \re{deltext}, and matched the soliton constants to small deviations from the
 corresponding stationary state. We considered the 2-soliton example \re{2nstxt} in detail and related its parameters to
 those of the asymptotic stationary state. Examples of $k=1,2$, and 3 normal soliton solutions
 are shown in Fig.~\ref{1nsfig}, \ref{2nsfig}, and \ref{3nsfig}. At large separation between the solitons, the $k$-soliton
 becomes a simple sum of $k$ single solitons, see e.g.  \eref{2nslargesep} and Figs.~\ref{2nsfig} and \ref{3nsfig}.

 Anomalous solitons asymptote to unstable eigenstates of the
 mean-field BCS Hamiltonian \re{anom} with nonzero anomalous average. We have obtained one, Eqs.~\re{deltaanomtxt},
 \re{indvsp}, and Fig.~\ref{1ansfig}, and two, Eqs.~\re{delans}, \re{f2an}, and Fig.~\ref{2asfig},  anomalous soliton solutions and
 related their parameters to those of
 the corresponding stationary states. The single soliton is a special case of a more general 3-spin solution, which we have
 also derived. In the vicinity of a stationary state, both normal and anomalous multi-solitons break up into a sum
 of single solitons. These single solitons are unstable normal modes in the linear analysis around the stationary state.

The utility of the soliton solutions  is that they are explicit and
are in terms of elementary functions (exponents), in contrast to the   general solution in terms of hyperelliptic
functions\cite{gensol}.
At the same time, the dynamics in many physical situations is multi-soliton. The combination of these two factors makes solitons
potentially quite useful in various problems in non-stationary superfluidity. Consider, for example, the collisionless dynamics
triggered  by an abrupt change of the pairing strength. In most cases of interest ${\bf L}^2(u)$
has only few isolated zeros, while the remaining complex zeros merge into continuous lines\cite{emilts}. We believe that the latter
zeros can be treated as  being degenerate   and their contribution is therefore multi-soliton. The solution  is then
 a superposition of a (quasi-)periodic few spin solution\cite{gensol,gensol1,dicke} with a multi-soliton
 one. Superpositions of this type are
 referred to as solitons on a (quasi-)periodic background in the soliton theory\cite{Kuznetsov}.
 In particular, when the system is in the ground
 state before the coupling change,
 the collisionless dynamics governed by \eref{bdg} can produce asymptotic states with a constant nonzero order
 parameter or  a gapless state\cite{emilts,Levitov2006,Emil2006}. In these cases
 ${\bf L}^2(u)$ has either a single pair of nondegenerate zeros
 or no such zeros. Therefore, according to the above reasoning, the dynamics
 leading to these
 asymptotic states is described by a multi-soliton solution of
 a normal type for the gapless state and of an anomalous type otherwise.

\section{Acknowledgements}

We thank M. Dzero for many stimulating discussions.
This research was financially supported by the National Science Foundation award NSF-DMR-0547769 and a David and Lucille Packard
Foundation Fellowship for Science and Engineering.


\begin{thebibliography}{99}
\bibitem{note1}   Indeed, the numerators of $L_{x}(u)$, $L_{y}(u)$, and $L_{z}(u)$ are polynomials in $u$ with real coefficients. Let $c$
be a real (double) zero of ${\bf L}^2(u)$ and let $a_x$, $a_y$, and $a_z$ be
the remnants from the division of these polynomials by $(u-c)$. Since $c$ is real
$a_{x,y,z}$ are also real. It follows from $L_x^2(u)+L_y^2(u)+L_z^2(u)={\bf L}^2(u)$ that $a_x^2+a_y^2+a_z^2=0$, i.e.
$a_x=a_y=a_z=0$.
\bibitem{note3} The frozen separation variables are the solutions of $L_s(u)=0$, where
$L_s(u)$ is given by \eref{ls}. Particle-hole symmetry implies $L_s(u)=-L_s(-u)$ and therefore  $\protect\sum_{j=3}^{n-1} u_j=0$.
\end{thebibliography}
\end{document}